\documentclass[journal,twocolumn,10pt]{IEEEtran}

\usepackage{cite}
\usepackage{amsmath,amssymb,amsfonts}
\usepackage{graphicx}
\usepackage{booktabs}
\usepackage{array}
\usepackage{siunitx}
\usepackage{url}
\usepackage{textcomp}
\usepackage{tikz}
\usepackage[hidelinks,hypertexnames=false]{hyperref}
\usepackage{algorithm}
\usepackage{algpseudocode}
\usepackage{subcaption}
\newlength{\scenePanelHeight}
\usepackage{booktabs}
\usepackage{tabularx}
\usepackage{array}

\newcolumntype{Y}{>{\raggedright\arraybackslash}X}

\usetikzlibrary{positioning,arrows.meta,shapes.geometric,calc,fit,backgrounds}

\makeatletter
\renewcommand{\ALG@name}{Algorithm}
\makeatother

\usepackage{fancyhdr}

\fancypagestyle{firstpageheader}{%
    \fancyhf{}
    \fancyhead[C]{%
        \parbox{0.95\textwidth}{%
            \centering
            \small
            Submitted for possible publication to IEEE.
            Paper currently under review.
            The contents of this paper may change at any time\\
            without notice.
        }%
    }

}

\begin{document}

\bstctlcite{IEEEexample:BSTcontrol}

\title{Calibrating the Digital Twin Channel: Statistics-Consistent Sim-to-Lab Adaptation for W-Band Industrial OFDM Links}

\author{Pulok Tarafder,~\IEEEmembership{Graduate Student Member,~IEEE}, Abigail O. Oyekola, Jasni Areepatta Mannil, Imtiaz Ahmed,~\IEEEmembership{Senior Member,~IEEE}, Zoheb Hassan,~\IEEEmembership{Member,~IEEE}, Danda B. Rawat,~\IEEEmembership{Senior Member},~IEEE, Wenjie Che,~\IEEEmembership{Member,~IEEE}

\thanks{P. Tarafder, A.O. Oyekola, J.A. Mannil, I. Ahmed, D. B. Rawat, and W. Che are with the Department of Electrical Engineering and Computer Science, Howard University, Washington, DC 20059, USA (emails: \{pulok.tarafder, abigail.oyekola, jasni.mannil\}@bison.howard.edu; imtiaz.ahmed@howard.edu;  danda.rawat@Howard.edu; wenjie.che@howard.edu).}
\thanks{Z. Hassan is with Dept. of Electrical and Computer Engineering, Université Laval, Québec, G1V 0A6, Canada (email: md-zoheb.hassan@gel.ulaval.ca).}

\thanks{This work was supported in part by the NSF Grant \# 2200640, in part by DoD/US Army Contract W911NF-22-1-022, and in part by the US DoD Center of Excellence in AI/ML at Howard University under Contract W911NF-20-2-0277 with the US ARL. However, the views and conclusions expressed herein are those of the authors and do not necessarily represent the official policies of the funding agencies.}}

\maketitle
\thispagestyle{firstpageheader}
\begin{abstract}
Digital twins (DTs) can reduce over-the-air validation cost in industrial wireless networks, but their utility depends on the fidelity of the underlying channel twin (CT). At W-band, site-specific ray tracing captures deterministic propagation geometry, yet its channel frequency responses (CFRs) do not reproduce the small-scale impairments and capture-to-capture variability observed in laboratory orthogonal frequency-division multiplexing (OFDM) measurements above 90~GHz. This paper proposes Statistics-Consistent Sim-to-Lab Adaptation (SC-SLA), a calibration framework that improves the fidelity of a 95~GHz Sionna ray-traced CT toward that of the testbed by aligning the mean power delay profile (PDP), the distribution of root-mean-square delay spread ($\tau_{\text{rms}}$), and per-subcarrier statistics at \SI{50}{\MHz} sampling bandwidth. SC-SLA uses a generative adversarial network (GAN)-inspired, cycle-consistent architecture with ResNet generators and batch-level channel-statistics losses on the PDP, sub-band PDP, $\tau_{\text{rms}}$ moments and quantiles, and normalized mean-square error (NMSE) of the mean CFR-magnitude profile. The framework is non-adversarial and requires neither paired simulated/measured samples nor discriminators. On held-out 95~GHz data, SC-SLA reduces the $\tau_{\text{rms}}$ Kolmogorov--Smirnov (KS) statistic from 0.86 to 0.050 relative to the impairment-augmented ray-traced input, and by $44\%$ ($0.089\!\to\!0.050$) relative to the strongest of four supervised baselines (FCNN, CNN1D, BiLSTM, and UNet1D). Without retraining, the same checkpoint also generalizes to 92--94~GHz carriers, where it achieves the lowest PDP and CFR-magnitude errors among all baselines while reducing the $\tau_{\text{rms}}$-distribution mismatch relative to the uncalibrated twin.
\end{abstract}

\begin{IEEEkeywords}
Digital twin channel, industrial IoT, W-band, ray tracing, OFDM, sim-to-real,
channel calibration, 6G.
\end{IEEEkeywords}

\section{Introduction}
\label{sec:intro}

\IEEEPARstart{I}{ndustry} 4.0 wireless deployments increasingly rely on dense collections of sensors, controllers, and mobile robots that require high-rate, low-latency, and reliable private connectivity~\cite{aceto2019survey, mahmood2022functional}. Validating every physical-layer configuration on hardware is costly, particularly in fixed industrial cells where machinery, racks, reflectors, and access points occupy known locations. This setting has motivated the use of \emph{digital twins} (DTs), measurement-synchronized virtual representations that can be queried before candidate configurations are transferred to the physical system~\cite{grieves2016dt,wu2021dtnsurvey,khan2022dt6g}. Industrial radio environments are well suited to this paradigm because their propagation geometry is largely static over the time scale of link-level design.

In a wireless DT, the channel determines link-level quantities such as throughput, pilot overhead, cyclic-prefix margin, and beam coherence. The \emph{digital twin channel}, referred to in this paper as the Channel Twin (CT), is therefore a foundation layer for wireless network-level twins~\cite{tarafder2026dtcorridor, wang2025dtc, tao2024wndt}. Under the high-frequency approximation of Maxwell's equations, a CT can model electromagnetic propagation through geometric rays and support geometry-aware analysis in complex three-dimensional environments~\cite{yun2015ray}. Site-specific ray tracing (RT) is a practical CT engine for industrial Internet of Things (IIoT) deployments. Once the cell geometry and material properties are specified, solvers such as NVIDIA Sionna RT~\cite{sionna, hoydis2023sionnart} and Wireless InSite~\cite{remcom_insite} can produce deterministic channel frequency responses (CFRs) and synthesize large datasets at low marginal cost. Recent advances further bring such solvers close to real-time operation~\cite{zhu2024raylaunch, alkhateeb2023rtdt}.

The usefulness of this closed loop depends on CT fidelity, which is difficult to guarantee in the frequency ranges targeted by 6G IIoT. The ITU-R IMT-2030 framework anticipates IMT operation in bands beyond those used for IMT-2020~\cite{itu2160}, and 3GPP has extended new radio (NR) operation beyond the original frequency range~2 (FR2) ceiling to 71~GHz~\cite{tr38808}. The W-band (75--110~GHz) follows this spectrum trajectory. In this paper, \emph{W-band} refers to the 92--95~GHz carriers under study. These carriers lie near the upper limit of TR~38.901~\cite{tr38901}, which is specified only up to 100~GHz and was parameterized primarily from sub-6~GHz and lower millimeter-wave (mmWave) measurements. Its small-scale channel assumptions therefore require measurement-based recalibration before they can support W-band physical-layer studies~\cite{wang2022thzsurvey, han2022sub}.

Calibration is precisely where ray-traced twins begin to fall short. In our comparison between 95~GHz Sionna CFRs and in-lab orthogonal frequency-division multiplexing (OFDM) measurements, two systematic fidelity gaps emerged. \textit{(i)~Early-tap energy underestimation.} The solver reproduces specular and refractive interactions but omits diffuse near-field scattering, antenna-to-cable mismatch ripple, and analog-front-end group-delay distortion, all of which contribute strongly to the first few channel-impulse-response (CIR) taps in a real W-band setup. \textit{(ii)~Narrow delay-spread distribution.} Because the geometry is static, the twin produces an almost deterministic root-mean-square (RMS) delay spread, $\tau_{\text{rms}}$, whereas the physical testbed exhibits a much broader capture-to-capture spread, attributable to effects such as oscillator phase noise, sampling-clock jitter, and signal-to-noise ratio (SNR) drift in the W-band frequency converter. A twin with these discrepancies provides only limited fidelity, and any IIoT link-level study built on it, including pilot-density selection, cyclic-prefix design, and beam-coherence analysis, inherits the resulting bias.

Learning the missing propagation and hardware effects from data is a natural calibration strategy. However, industrial measurements impose an important constraint. Supervised calibration would require paired simulated and measured CFRs for the same channel realization, which is rarely available in field or laboratory captures. Simulated and measured acquisitions are typically generated independently, leaving no sample-wise correspondence between domains. The calibration problem must therefore be treated as an \emph{unpaired} domain-alignment task. The learned correction should also preserve channel statistics that affect OFDM receiver design, including the power delay profile (PDP), $\tau_{\text{rms}}$, and per-subcarrier magnitude, rather than merely fitting individual waveform samples.

Motivated by these requirements, we propose Statistics-Consistent Sim-to-Lab Adaptation (SC-SLA), a calibration layer between a ray-traced CT and a physical W-band testbed. SC-SLA learns an unpaired, non-adversarial mapping from Sionna-generated CFRs to testbed-like CFRs by directly aligning receiver-relevant channel statistics. The calibrated twin retains the scalability of RT while reproducing the delay and frequency-domain statistics observed in the measured industrial link.

\subsection{Related Work}
\label{sec:related}

Outside explicit wireless-DT calibration, prior efforts to reduce
the gap between modeled and measured wireless channels have followed
three main directions. \emph{Measurement-based and site-specific
models} extend geometry-based stochastic modeling beyond mmWave by
incorporating empirical path-loss, clustering, and delay
statistics~\cite{rappa2019,ju2021jsac}. Recent studies also infer
channel statistics from environment geometry or co-located sensor
data~\cite{song2026sitespecific,mi2024pointcloud}. These models capture
propagation behavior. However, they do not calibrate the capture-specific
impairments introduced by a particular measurement testbed.

\emph{Supervised networks trained on measured channels} have been used
for channel prediction and estimation, including vehicular channel
state information (CSI) prediction~\cite{joo2019vehicular} and massive
multiple-input multiple-output (MIMO) channel prediction validated on
field measurements~\cite{shehzad2022realworld}. In sim-to-real
calibration, however, these methods require synchronized simulated and
measured channel pairs, which are rarely available in field or
industrial captures.

\emph{Generative models} relax the need for paired samples by learning
the underlying channel distribution. Examples include generative
adversarial networks (GANs) trained on measured channel-sounder
data~\cite{euchner2024ganmimo}, conditional GANs for air-to-ground
channels~\cite{tian2024a2g}, and physics-informed generative
variants~\cite{bock2025physinf}. Nevertheless, these approaches
typically optimize adversarial or parameter-domain objectives without
explicitly preserving OFDM-relevant statistics such as PDP shape, $\tau_{\text{rms}}$, and frequency-domain magnitude structure.

A parallel line of work focuses explicitly on reducing the sim-to-real
gap in wireless DTs. Following the taxonomy in
\cite{ruah2025bridge}, this gap can be addressed through
(i) direct calibration of the DT using real measurements,
(ii) uncertainty-aware modeling of the residual environment mismatch,
and (iii) correction of the task-level AI training objective. SC-SLA
belongs to the first category and, more specifically, performs
\emph{post-RT channel-output calibration}: the RT solver, reconstructed
geometry, and nominal material properties remain fixed while a learned
mapping aligns the simulated and measured CFR distributions.

Direct DT calibration can operate at different stages of the
channel-generation process. Model-space approaches learn object-level
electromagnetic properties and radio-wave interaction
functions~\cite{jiang2025learnable}, or estimate RT material parameters
while compensating for path-phase errors caused by geometric
mismatch~\cite{ruah2024phase}. Geometry-space calibration instead
corrects physical scene variables, such as transmitter and receiver
locations, using measured and simulated PDPs~\cite{ying2026nyuray}.
Propagation-consistent environment twins can also embed a learnable
scene-level electromagnetic field within differentiable RT and
calibrate it using sparse position-labeled CSI~\cite{ai2026wedt}.

Channel-output calibration avoids reconstructing the complete internal
environment model. Haider et al.~\cite{haider2025dtprecoding} first
increase the fidelity of a site-specific RT model by selecting suitable
propagation settings and material assignments, and then train a
supervised U-Net to refine the resulting RT CIRs using measured
over-the-air CIRs as labels. The refined channels are subsequently used
for precoding and evaluated through end-to-end bit error rate (BER). Similarly,
Luo et al.~\cite{luo2026dft} use supervised learning to refine
discrete Fourier transform (DFT)-domain channel weights generated by a
low-complexity DT for CSI
compression and feedback. Table~\ref{tab:dt_calibration_comparison}
summarizes these calibration approaches and positions SC-SLA relative
to them.

  \begin{table*}[t]
  \caption{Representative wireless-DT calibration approaches}
  \label{tab:dt_calibration_comparison}
  \centering
  \scriptsize
  \setlength{\tabcolsep}{3pt}
  \renewcommand{\arraystretch}{1.12}
  \begin{tabularx}{\textwidth}{
  @{}
  >{\raggedright\arraybackslash}p{1.85cm}
  >{\raggedright\arraybackslash}p{1.35cm}
  >{\raggedright\arraybackslash}p{2.15cm}
  >{\raggedright\arraybackslash}p{2.45cm}
  >{\raggedright\arraybackslash}p{2.60cm}
  Y
  @{}}
  \toprule
  \textbf{Work}
  &
  \textbf{Calibration space}
  &
  \textbf{Calibrated component}
  &
  \textbf{Measurement or supervision}
  &
  \textbf{Calibration objective}
  &
  \textbf{Relation to SC-SLA}
  \\
  \midrule

  Jiang~et~al.~\cite{jiang2025learnable}
  &
  Model
  &
  Object-level electromagnetic representation and interaction model
  &
  Scene-associated wireless observations
  &
  Learn electromagnetic properties and radio-wave interactions
  &
  Re-learns the environment model. SC-SLA keeps the RT model fixed and
  calibrates its CFR outputs.
  \\
  \addlinespace[2pt]

  Ruah~et~al.~\cite{ruah2024phase}
  &
  Model
  &
  RT material parameters and path-phase errors
  &
  Position-associated channel observations
  &
  Estimate material parameters while accounting for geometric phase
  errors
  &
  Calibrates inside a differentiable RT solver. SC-SLA does not
  backpropagate through the solver.
  \\
  \addlinespace[2pt]

  Ying~et~al.~\cite{ying2026nyuray}
  &
  Geometry
  &
  Transmitter and receiver locations
  &
  Measured and simulated PDPs at corresponding sites
  &
  Correct location uncertainty and multipath mismatch
  &
  Corrects TX/RX placement. SC-SLA assumes fixed nominal node positions.
  \\
  \addlinespace[2pt]

  Ai~et~al.~\cite{ai2026wedt}
  &
  Environment
  &
  Scene-level electromagnetic property field
  &
  Sparse position-labeled CSI
  &
  Construct a propagation-consistent environment representation
  &
  Modifies the environment model and requires differentiable RT. SC-SLA
  applies a post-RT CFR mapping.
  \\
  \addlinespace[2pt]

  Haider~et~al.~\cite{haider2025dtprecoding}
  &
  Output
  &
  RT configuration, material assignments, and generated CIR
  &
  Paired RT-generated and measured CIRs
  &
  Refine CIRs using a supervised U-Net for precoding and BER evaluation
  &
  Uses paired CIR supervision and task-level validation. SC-SLA uses
  unpaired data and ensemble statistics.
  \\
  \addlinespace[2pt]

  Luo~et~al.~\cite{luo2026dft}
  &
  Output
  &
  DFT-domain channel output
  &
  Aligned low-/high-fidelity channel information or historical CSI
  &
  Refine DFT-domain channel weights by supervised learning
  &
  Uses an aligned, task-oriented representation. SC-SLA calibrates
  unpaired complex OFDM CFRs.
  \\
  \addlinespace[2pt]

  \textbf{SC-SLA}
  &
  \textbf{Output}
  &
  \textbf{Post-RT complex CFR distribution}
  &
  \textbf{Independently acquired Sionna RT and W-band testbed captures}
  &
  \textbf{Align PDP, sub-band PDP, $\tau_{\text{rms}}$, and
  per-subcarrier magnitude statistics}
  &
  \textbf{Unpaired, non-adversarial calibration that leaves the RT scene,
  solver, and communication model unchanged.}
  \\

  \bottomrule
  \end{tabularx}
  \end{table*}

Adjacent sim-to-real studies adapt channel generators or communication
models using limited measurements. Hu et al.~\cite{hu2024ttgan}
propose a transfer-learning transformer GAN pretrained on THz channels
from a geometry-based stochastic channel model and fine-tuned using a
smaller vector network analyzer measurement set. The model remains
adversarial and operates in the multipath-parameter domain.
Baytekin et al.~\cite{baytekin2026nrx} instead pretrain a neural
receiver on a randomized 3GPP urban-microcell channel model and
fine-tune it with real 5G NR physical uplink shared channel (PUSCH)
measurements. Their objective is
end-to-end block-error-rate performance rather than calibration of the
underlying channel distribution.

Taken together, prior approaches leave three requirements unresolved for the present setting: (i) calibrating the generated channel without modifying or differentiating through the RT model, (ii) learning from
independently acquired simulated and measured CFRs without sample-wise correspondence, and (iii) correcting the channel
distribution itself rather than optimizing a task-specific communication objective. Fidelity calibration of a ray-traced CT for a single-antenna W-band OFDM link therefore remains largely unaddressed, particularly under the unpaired acquisition conditions considered in this work.

\subsection{Contributions}
\label{sec:contrib}

The proposed SC-SLA addresses this gap by providing an unpaired, GAN-inspired but non-adversarial calibration layer for the CT of a static industrial communication network. Instead of requiring sample-wise simulated/measured correspondence, SC-SLA directly aligns measured channel statistics that govern OFDM link behavior. The main contributions are summarized as follows.

\begin{itemize}

\item \textbf{An unpaired distributional CT-calibration formulation:}
We formulate sim-to-lab adaptation as the problem of bringing the
statistical distribution of DT-generated CSI toward that of
independently acquired real-world industrial CSI captures. The formulation explicitly
aligns complementary delay and frequency-domain attributes, including
the mean PDP, local sub-band delay structure, the
distribution of $\tau_{\text{rms}}$, and the per-subcarrier magnitude
profile, without imposing an arbitrary one-to-one correspondence
between channel realizations.

\item \textbf{A W-band OFDM testbed for CT calibration:}
We develop a commercial off-the-shelf (COTS) W-band measurement platform that
combines USRP-B200 radios, WR-10 frequency-conversion modules, Faraday
isolators, and pyramidal horn antennas. The testbed supports complex
OFDM CFR acquisition over 92--95~GHz on the same subcarrier grid used
by the ray-traced CT, thereby providing the real-world CSI required for
sim-to-lab calibration and cross-frequency evaluation.

\item \textbf{A CycleGAN-style non-adversarial channel translator:} We develop a two-generator, cycle-consistent translation architecture inspired by GAN-based unpaired domain adaptation, but replace adversarial discriminators with receiver-relevant channel-statistics losses. The training objective combines $\ell_1$ losses on the truncated PDP and ten sub-band PDPs, $\tau_{\text{rms}}$ moment and quantile losses, normalized mean-square error (NMSE) of the mean CFR-magnitude profile, and cycle/identity regularization.

\item \textbf{An empirical study at 92--95~GHz:}
We evaluate SC-SLA using Sionna~RT CFRs from the reconstructed
laboratory CT and OFDM CFR measurements from the W-band testbed. At
95~GHz, SC-SLA reduces the $\tau_{\text{rms}}$
Kolmogorov--Smirnov (KS) statistic by $44\%$ relative to the strongest of
four supervised baselines while also achieving the lowest PDP and
CFR-magnitude NMSE. Without retraining or carrier-specific tuning, the
same checkpoint transfers to 92, 93, and 94~GHz and retains the lowest
PDP and CFR-magnitude errors among the evaluated baselines, demonstrating
calibration robustness across a 3~GHz carrier shift. To support
reproducibility and further research on CT calibration, we publicly
release the Sionna datasets, testbed datasets, trained checkpoints, and
source code\footnote{Available at
\url{https://github.com/puloktarafder/thz-scsla} upon publication.}.

\end{itemize}

The remainder of this paper is organized as follows.
Section~\ref{sec:sysmodel} presents the signal model and the simulated and measured datasets. Section~\ref{sec:method} details the SC-SLA generator and its channel-statistics losses.
Section~\ref{sec:experiments} reports in-domain and cross-frequency results against the supervised baselines, together with inference scalability, sensitivity to the amount of measured training data, and an ablation of the loss terms and generator head that isolates the mechanisms behind the observed gains.
Section~\ref{sec:conclusion} concludes the paper.
\section{System Model and Experimental Setup}
\label{sec:sysmodel}

This section specifies the OFDM signal model, the simulated and measured CFR datasets, and the common normalization protocol used by SC-SLA. The simulated ray-traced CT dataset is denoted by $X_s$, and its measured testbed counterpart is denoted by $X_r$. Both datasets use the same OFDM grid and normalization procedure before calibration.
\begin{figure}[t]
\centering
\begin{subfigure}[t]{0.48\linewidth}
\centering
\includegraphics[width=\linewidth]{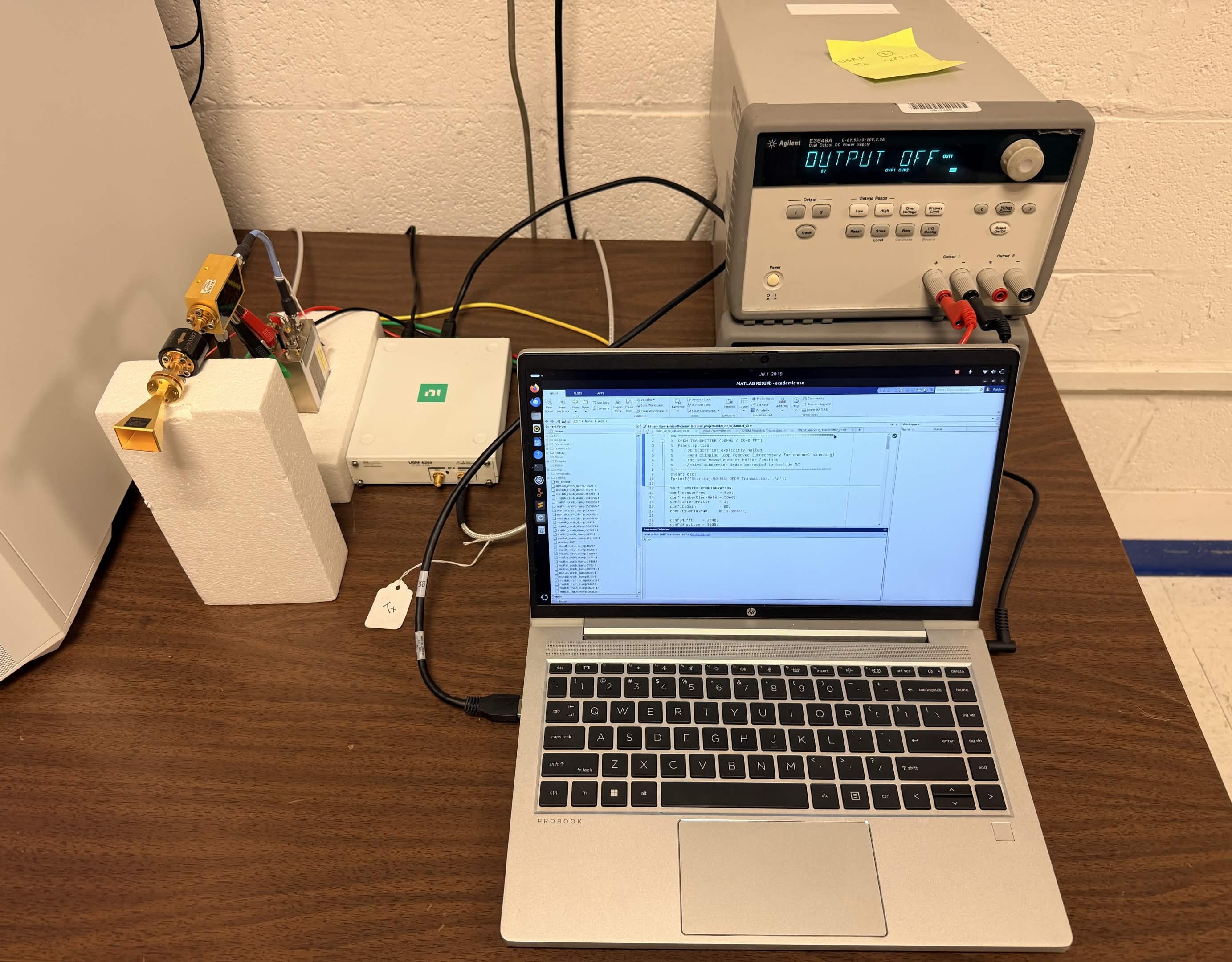}
\caption{Transmitter side.}
\label{fig:testbed-tx}
\end{subfigure}
\hfill
\begin{subfigure}[t]{0.48\linewidth}
\centering
\includegraphics[width=\linewidth]{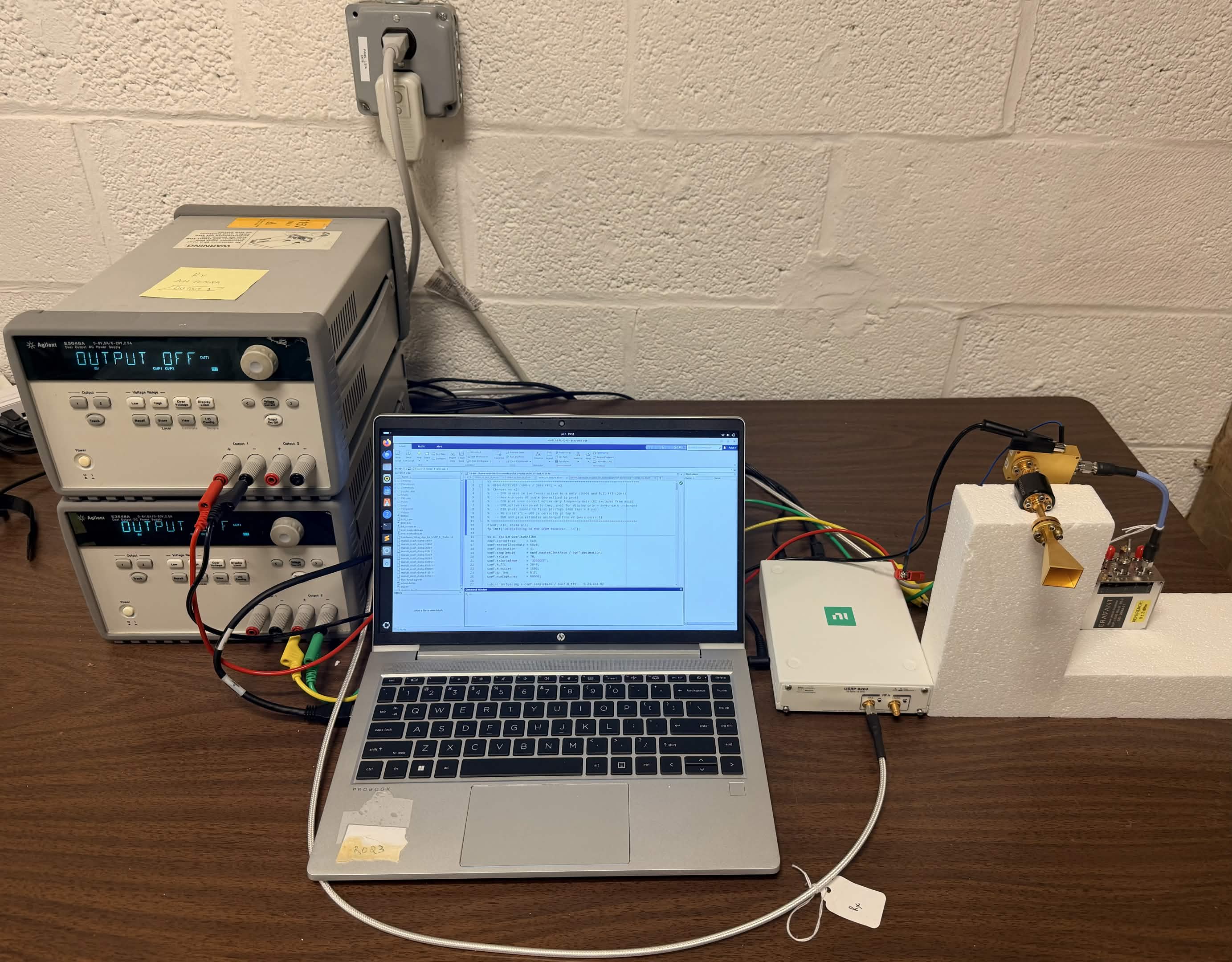}
\caption{Receiver side.}
\label{fig:testbed-rx}
\end{subfigure}
\caption{W-band OFDM testbed setup. The transmitter and receiver sides use Eravant WR-10 pyramidal horn antennas that are co-polarized and boresight-aligned to form the LOS measurement link.}
\label{fig:testbed}
\end{figure}

\begin{figure*}[ht!]
  \centering
  \setlength{\scenePanelHeight}{0.195\linewidth}
    \begin{subfigure}[t]{0.36\linewidth}
      \centering
      \includegraphics[height=\scenePanelHeight]{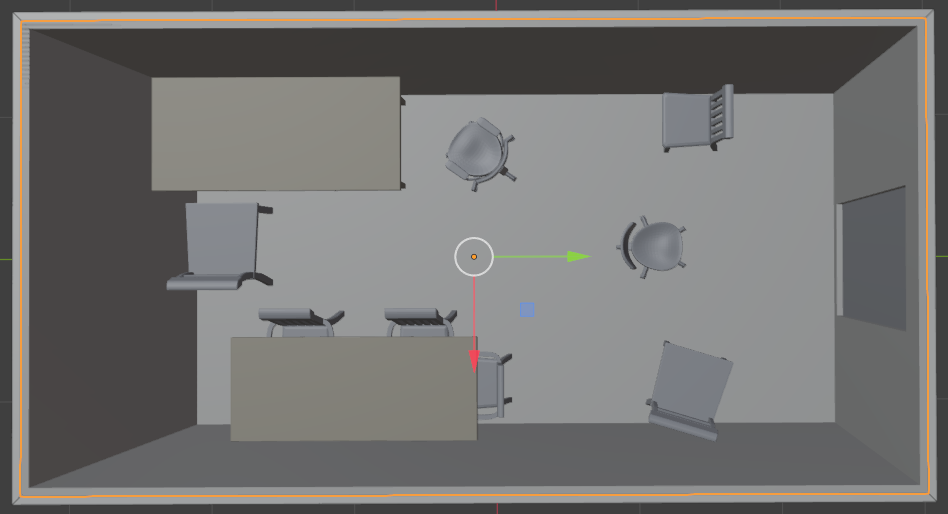}
      \caption{Blender scene.}
      \label{fig:scene-blender}
    \end{subfigure}\hfill
    \begin{subfigure}[t]{0.295\linewidth}
      \centering
      \includegraphics[height=\scenePanelHeight]{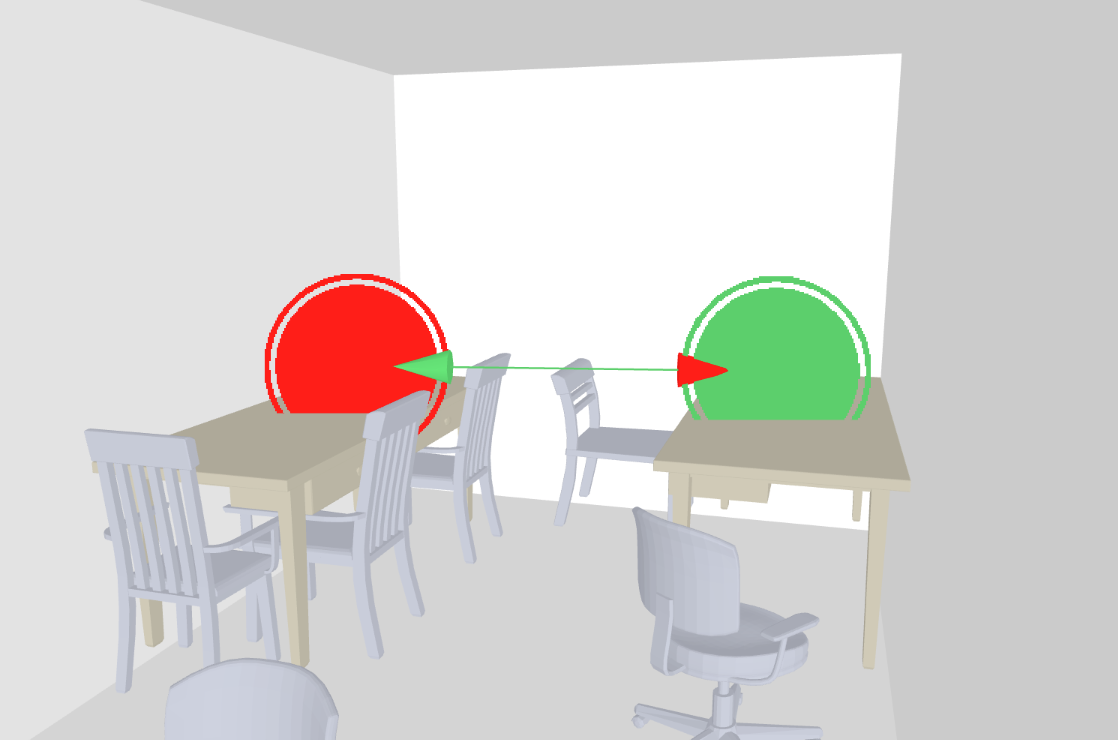}
      \caption{Sionna scene.}
      \label{fig:scene-sionna}
    \end{subfigure}
    \begin{subfigure}[t]{0.295\linewidth}
      \centering
      \includegraphics[height=\scenePanelHeight]{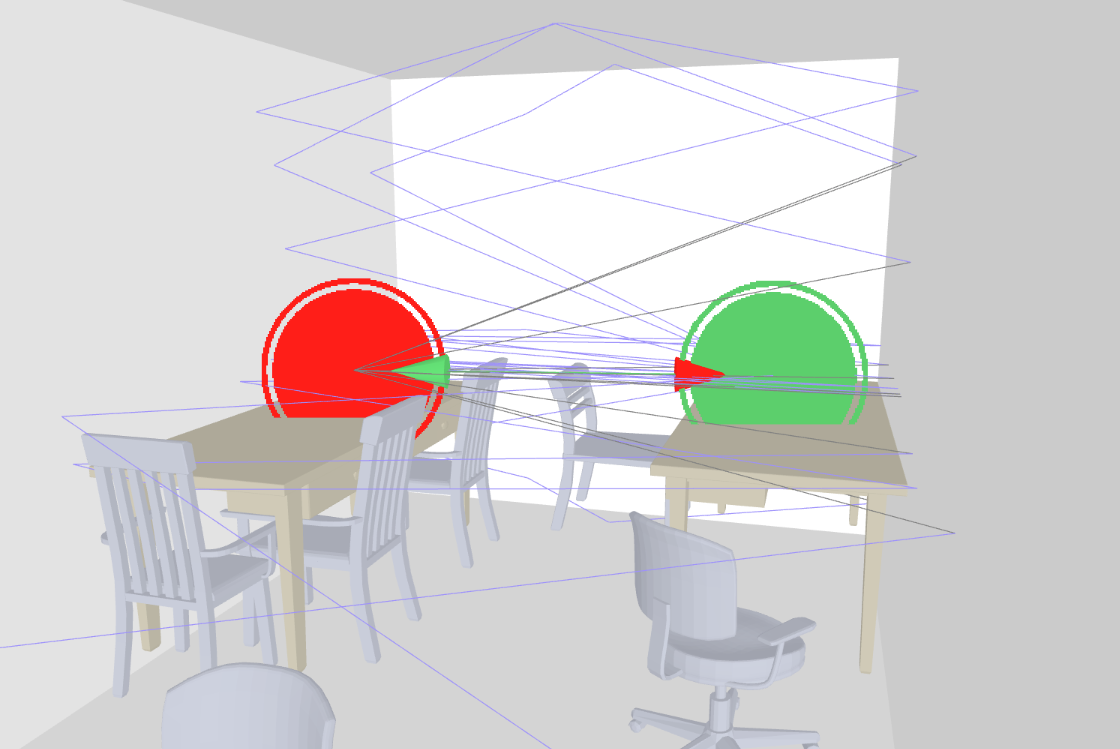}
      \caption{Sionna RT scene.}
      \label{fig:scene-rt}
    \end{subfigure}\hfill
  \caption{Ray-traced CT construction. (a) Blender reconstruction of the indoor W-band testbed. (b) Imported Sionna~RT scene with the transmitter and receiver placed at the nominal horn antenna locations. (c) Ray-traced Sionna~RT scene with LOS, specular reflection, and refraction paths overlaid.}
  \label{fig:rt_scenes}
\end{figure*}

\subsection{OFDM Signal and Channel Model}
\label{sec:signal}
We consider a static indoor W-band link over 92--95~GHz with a dominant line-of-sight (LOS) component. To make simulation and measurement directly comparable, both domains use the same OFDM reference grid. Each OFDM symbol has $N_{\text{FFT}}=2048$ subcarriers, of which $N_{\text{a}}=1500$ are active. The active subcarriers occupy
\[\mathcal{K}_{\text{a}} = \{+1,\dots,+750\}\cup\{-750,\dots,-1\}, \]
which excludes the direct-current (DC) bin $k=0$. This bin is left unused because practical radio-frequency (RF) front ends can exhibit DC offset, local-oscillator (LO) leakage, mixer feedthrough, and residual in-phase/quadrature (I/Q) imbalance around the carrier. Excluding the DC tone therefore avoids a hardware-sensitive subcarrier and yields a cleaner active CFR. The sampling bandwidth is $B_{\text{s}}=\SI{50}{\MHz}$, yielding a subcarrier spacing of $\Delta f=B_{\text{s}}/N_{\text{FFT}}\approx\SI{24.41}{\kHz}$ and a useful OFDM
symbol duration of $T_u=1/\Delta f\approx\SI{40.96}{\micro\second}$. The active subcarriers therefore span approximately
$N_{\text{a}}\Delta f = 1500 \times \SI{24.41}{\kHz} \approx \SI{36.6}{\MHz}$ of the \SI{50}{\MHz} sampling bandwidth, with the remainder serving as guard band. A cyclic prefix of $N_{\text{cp}}=512$ samples is prepended to each symbol. The active CFR on subcarrier $k\in\mathcal{K}_{\text{a}}$ is modeled as
\begin{equation}
H[k] = \sum_{\ell=0}^{L-1} \alpha_\ell\, e^{-j2\pi k \Delta f \tau_\ell} + Z[k],
\label{eq:cfr}
\end{equation}
where $\alpha_\ell$ and $\tau_\ell$ denote the complex gain and excess delay of path $\ell$, respectively, $L$ is the number of resolvable paths, and $Z[k]$ represents the residual receiver noise after least-squares (LS) channel estimation.

Let $\tilde H[m]$ denote the full $N_{\text{FFT}}$-point CFR obtained by placing the active samples $H[k]$ on their corresponding fast Fourier transform (FFT) bins and assigning zeros to the DC and guard-band bins. The truncated CIR is obtained from the $N_{\text{FFT}}$-point inverse FFT (IFFT) of $\tilde H[m]$ by retaining the first $T=20$ taps,
\begin{equation}
\tilde h[n] = \frac{1}{N_{\text{FFT}}}\sum_{m=0}^{N_{\text{FFT}}-1}
\tilde H[m]\,e^{j2\pi m n/N_{\text{FFT}}},\quad n=0,\dots,T-1.
\label{eq:cir}
\end{equation}
Because the IFFT is defined over the full sampling bandwidth $B_{\text{s}}$, adjacent CIR taps are spaced by $\Delta t = 1/B_{\text{s}} = \SI{20}{\nano\second}$. Retaining $T=20$ taps gives a uniformly sampled \SI{400}{\nano\second} delay window. From this truncated CIR, the normalized PDP is computed as $p[n] = |\tilde h[n]|^2/\sum_{m=0}^{T-1}|\tilde h[m]|^2$. The resulting discrete delay-energy distribution defines $\tau_{\text{rms}}$ as
\begin{equation}
\tau_{\text{rms}} = \sqrt{\sum_{n=0}^{T-1}\!(n\Delta t)^2 p[n] -
\!\Big(\sum_{n=0}^{T-1}\!n\Delta t\, p[n]\Big)^{2}}. \label{eq:taurms}
\end{equation}
The finite delay support reduces the influence of late-delay noise-floor samples on the second-order delay moment. This choice is consistent with prior observations that noise and spurious PDP components at large delays can bias measured $\tau_{\text{rms}}$ estimates if they are not excluded before computation~\cite{schulpen2023ambiguity}. The CFR, truncated CIR, PDP, and $\tau_{\text{rms}}$ definitions above provide the common channel representation used to construct $X_s$ and $X_r$.

\subsection{Physical Testbed (W-Band USRP Measurements)}
\label{sec:realdata}
The measured dataset is captured with the W-band OFDM testbed in Fig.~\ref{fig:testbed}, which operates over 92--95~GHz and uses COTS components. Separate USRP-B200 devices serve as transmitter (TX) and receiver (RX). The transmitter USRP generates the baseband OFDM waveform, while the receiver USRP records the received waveform with a \SI{50}{\MHz} master clock and a \SI{70}{\dB} receiver-gain setting. The waveform is generated at an intermediate frequency (IF), $f_{\text{IF}}\in\{2,3,4,5\}~\si{\GHz}$, upconverted to W-band at the transmitter, and downconverted back to IF at the receiver using matched COTS Eravant~\cite{eravant2026} WR-10 up/down-conversion modules. The chain uses a $\times 8$-multiplied phase-locked oscillator and yields $f_{\text{RF}} = 8f_{\text{LO}} + f_{\text{IF}}$, where $f_{\text{LO}}=11.25$~GHz. Retuning $f_{\text{IF}}$ produces carriers from $92$ to $95$~GHz. A W-band Faraday isolator is inserted at each antenna port using COTS WR-10 waveguide components operating over 75--110~GHz, with \SI{28}{\dB} isolation, \SI{4.5}{\dB} insertion loss, and a \SI{90}{\degree} twist. Both ends use co-polarized, boresight-aligned Eravant WR-10 pyramidal horn antennas, with a TX--RX separation of $\SI{1.75}{\metre}$.

The transmitted reference is a flat-power OFDM symbol whose active subcarriers carry unit-magnitude symbols with uniformly random phases. For each carrier, the receiver records a \SI{15}{\second} burst. Offline synchronization cross-correlates the received burst with the known reference symbol. Candidate frame peaks are detected with a minimum peak distance of $N_{\text{FFT}}+N_{\text{cp}}$ and a threshold equal to three times the mean correlation magnitude. Each valid detected reference symbol yields one channel capture. For every detected capture, the receiver computes the FFT of the synchronized OFDM symbol and estimates the active CFR by the LS method, $\hat H[k]=\frac{Y[k]}{X_{\text{ref}}[k]}$, where $Y[k]$ is the received pilot on active subcarrier $k$ and $X_{\text{ref}}[k]$ is the transmitted reference symbol. The estimated active CFR is stored over the same $1500$ active tones and in the same $[+1,\dots,+750,-750,\dots,-1]$ ordering used by the simulated dataset. A full $N_{\text{FFT}}$-point CFR is also constructed by inserting zeros on the DC and guard-band bins, and its IFFT gives the corresponding full-length CIR.

An instantaneous SNR is estimated for each capture from the full CIR. The signal power is taken as the maximum tap power within the first $100$ taps, and the noise power is estimated as the mean tap power over the remaining taps. Captures with SNR below \SI{8}{\dB} are rejected. After SNR filtering, we retain $N_r=50{,}000$ valid CFRs per carrier. At 95~GHz, the retained captures have a mean estimated SNR of \SI{28.98}{\dB} and a standard deviation of \SI{2.87}{\dB}. For 95~GHz training, the measured dataset is split into $45{,}000$ training samples and $5{,}000$ held-out samples, and $2{,}000$ held-out captures are subsampled for each reported metric.

\begin{figure*}[t]
\centering
\includegraphics[width=\textwidth]{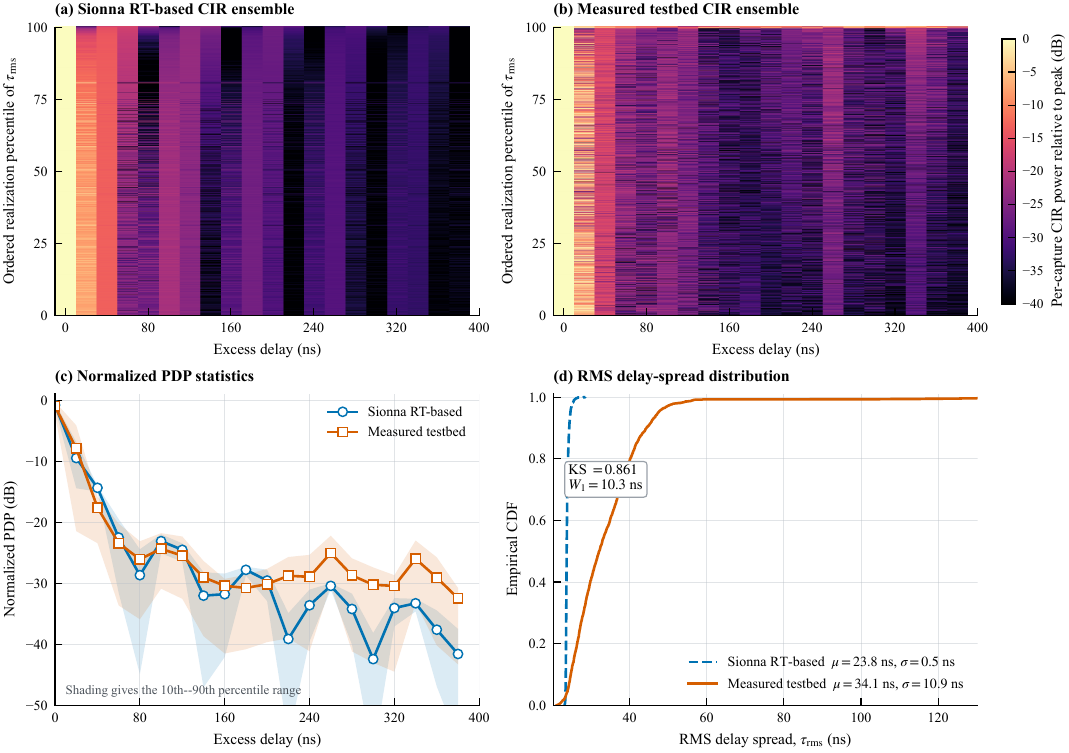}
\caption{Delay-domain comparison of the impairment-augmented Sionna RT and measured testbed channels at 95~GHz using $2{,}000$ unpaired held-out samples per domain. Panels (a) and (b) depict the first 20 CIR taps, normalized to each sample's peak power and ordered by $\tau_{\text{rms}}$. Panel (c) presents the mean unit-energy PDP and its 10th--90th percentile range. Panel (d) compares the empirical cumulative distribution functions and reports the mean and standard deviation of $\tau_{\text{rms}}$ as $(\mu,\sigma)=(23.8,0.5)$~\si{\nano\second} for Sionna RT and $(34.1,10.9)$~\si{\nano\second} for the measured testbed, together with the KS statistic $D_{\mathrm{KS}}=0.861$ and the empirical 1-Wasserstein distance $W_1=\SI{10.3}{\nano\second}$.}
\label{fig:cir_statistics_gap}
\end{figure*}

The retained measurements exhibit substantial delay-domain variation even though the antenna positions and link geometry remain fixed. Across the $2{,}000$ held-out 95~GHz captures in Fig.~\ref{fig:cir_statistics_gap}, the mean $\tau_{\text{rms}}$ is \SI{34.05}{\nano\second}, with a standard deviation of \SI{10.88}{\nano\second}. Its 10th and 90th percentiles are \SI{24.75}{\nano\second} and \SI{44.40}{\nano\second}, respectively. The normalized CIR and PDP statistics indicate that the variation redistributes energy among the retained taps rather than producing only a common gain change. These measurements define the reference distribution used to assess RT fidelity.

\subsection{Ray-Traced CT (Sionna RT)}
\label{sec:simdata}
The simulated CT dataset is generated with the Sionna~RT path solver~\cite{hoydis2023sionnart} on a calibrated indoor scene. The geometry corresponds to a single-room laboratory reconstructed in Blender, with walls and floor assigned ITU-R indoor material properties at W-band. The scene is exported as a Mitsuba XML file. Fig.~\ref{fig:rt_scenes} depicts the Blender reconstruction, the imported Sionna scene, and the traced paths. The TX and RX are placed at the nominal horn locations with a LOS distance of $\SI{1.75}{\metre}$, matching the testbed. At both ends, we use a custom vertically polarized antenna element with a horn-like directive pattern, since Sionna~RT does not provide a native horn model. The antenna power pattern is modeled as $P_{\mathrm{h}}(\vartheta)\propto\max(\cos\vartheta,0)^{n_{\mathrm{h}}}$, where $\vartheta$ is the angular offset from boresight. The exponent $n_{\mathrm{h}}=\ln(1/2)/\ln[\cos(6^\circ)]\approx126.18$ solves the half-power condition at a $6^\circ$ half-angle, giving a \SI{12}{\degree} full 3-dB beamwidth. The cosine-power model captures the dominant boresight directivity while omitting detailed sidelobe structure. We configure the path solver to support LOS, specular reflection, and refraction with a maximum interaction depth of $5$ and disable diffuse reflection. The resulting RT model represents the deterministic geometric component of the link, while PC-SLA learns the residual, comprising both stochastic capture variation and the deterministic response of the measurement chain.
For each carrier, the static scene is solved once to obtain a deterministic base CFR over the same active OFDM tones used by the measurement waveform. The active-frequency vector follows the ordering $[+1,\dots,+750,-750,\dots,-1]$,
which matches the stored active-bin convention of the measured CFRs. The path delays are normalized so that the first arrival falls at tap~0, matching the synchronization convention used in the receiver processing.

A single RT solution yields one deterministic base CFR per carrier, whereas the physical testbed produces a distribution of captures because each acquisition is affected by noise, timing, and phase fluctuations. To emulate this behavior and match the $N_r$ captures retained per carrier in Section~\ref{sec:realdata}, we generate $N_s=50{,}000$ simulated samples per carrier by passing the deterministic base CFR through a measurement-inspired impairment layer. Let $H_{\mathrm{RT}}[k]$ denote the deterministic base CFR on active subcarrier $k$, and let $i$ index the simulated capture. For each capture, a timing offset $\Delta\tau_i$ is drawn independently from a zero-mean Gaussian distribution with a standard deviation of \SI{2.5}{\nano\second}. This offset is applied as the subcarrier-dependent phase rotation $e^{-j2\pi k\Delta f\,\Delta\tau_i}$, which introduces a random linear phase slope across frequency. A common phase offset $\phi_i$ is independently drawn from a zero-mean Gaussian distribution with a standard deviation of \SI{0.20}{\radian} and applied identically to all active subcarriers. The resulting impairment-augmented ray-traced CFR is
\begin{equation}
\begin{aligned}
H_i^{(s)}[k]
&=
H_{\mathrm{RT}}[k]\,
e^{\,j\phi_i-j2\pi k\Delta f\,\Delta\tau_i}
+V_i[k],
\end{aligned}
\label{eq:simulated_impairments}
\end{equation}
where $V_i[k]$ is zero-mean complex Gaussian noise scaled from the mean base-CFR power to realize a per-capture SNR drawn from a Gaussian distribution with mean \SI{28}{\dB} and standard deviation \SI{2}{\dB}, limited to 8--60~dB. These nominal levels provide conservative capture variation without tuning to the held-out $\tau_{\text{rms}}$ distribution. The \SI{2.5}{\nano\second} timing spread equals one eighth of the \SI{20}{\nano\second} sampling interval, and the \SI{0.20}{\radian} common-phase spread is about \SI{11.5}{\degree}. These phase perturbations leave the path geometry unchanged. The simulated SNR mean and standard deviation, \SI{28}{\dB} and \SI{2}{\dB}, approximate the corresponding measured values of \SI{28.98}{\dB} and \SI{2.87}{\dB} for the retained 95~GHz captures. The \SI{8}{\dB} lower limit matches the measurement rejection threshold, while the \SI{60}{\dB} upper limit suppresses extreme Gaussian draws. Further details on the typical Sionna~RT workflow and site-specific dataset construction are available in our prior works~\cite{tarafder2025digital,tarafder2026dtcorridor}.

Fig.~\ref{fig:cir_statistics_gap} summarizes the residual delay-domain gap after capture-side impairment augmentation. Across the $2{,}000$ held-out samples from each domain, the Sionna RT channels have a mean $\tau_{\text{rms}}$ of \SI{23.82}{\nano\second} and a standard deviation of \SI{0.46}{\nano\second}, compared with \SI{34.05}{\nano\second} and \SI{10.88}{\nano\second} for the measured channels. The measured standard deviation is $23.6\times$ larger. Let $\tau^{(s)}_{(i)}$ and $\tau^{(r)}_{(i)}$ denote the $i$th ordered $\tau_{\text{rms}}$ samples from the simulated and measured domains, respectively. For $N=2{,}000$ equally weighted samples per domain, the empirical one-dimensional 1-Wasserstein distance is~\cite{villani2009}
\begin{equation}
W_1=\frac{1}{N}\sum_{i=1}^{N}
\left|\tau^{(s)}_{(i)}-\tau^{(r)}_{(i)}\right|.
\label{eq:wasserstein1}
\end{equation}
The two distributions yield a two-sample KS statistic of $D_{\mathrm{KS}}=0.861$, defined in Eq.~\eqref{eq:ks}, and $W_1=\SI{10.3}{\nano\second}$. These distributional discrepancies demonstrate that nominal capture-side impairments do not reproduce the measured delay-domain variability and motivate the learned SC-SLA calibration.

\subsection{Joint Normalization}
\label{sec:norm}
The simulated and measured datasets pass through a common normalization procedure before training and evaluation. The goal is to remove absolute-power and receiver-gain differences while preserving the relative phase, frequency selectivity, and delay-domain structure needed for channel-statistics alignment.

First, each complex CFR sample is divided by its own RMS magnitude. This per-sample normalization removes capture-to-capture gain variation without forcing the two datasets to share a global power scale. Second, the normalized complex CFR is converted into a real-valued tensor by interleaving the real and imaginary components along the channel dimension, yielding samples of size $(N_{\text{a}},2)$ and dataset tensors of shape $(N_s,N_{\text{a}},2)$ and $(N_r,N_{\text{a}},2)$ for the simulated and measured sets, respectively. Third, let $\rho_{\text{clip}}=0.995$ denote the joint quantile level. The threshold $q_{\text{clip}}$ is the $\rho_{\text{clip}}$-quantile of the pooled simulated and measured magnitudes. Both datasets are clipped at $q_{\text{clip}}$ and scaled to the normalized support expected by the generator.

The normalization parameters are fixed after training and reused during evaluation. This keeps the input normalization identical across carriers, allowing the $95$~GHz checkpoint to be applied directly to the $92$--$94$~GHz carriers without retuning. All SC-SLA losses and reported metrics are computed in this shared normalized representation.

\section{SC-SLA Framework}
\label{sec:method}
\begin{figure*}[t]
\centering
\includegraphics[width=0.95\textwidth]{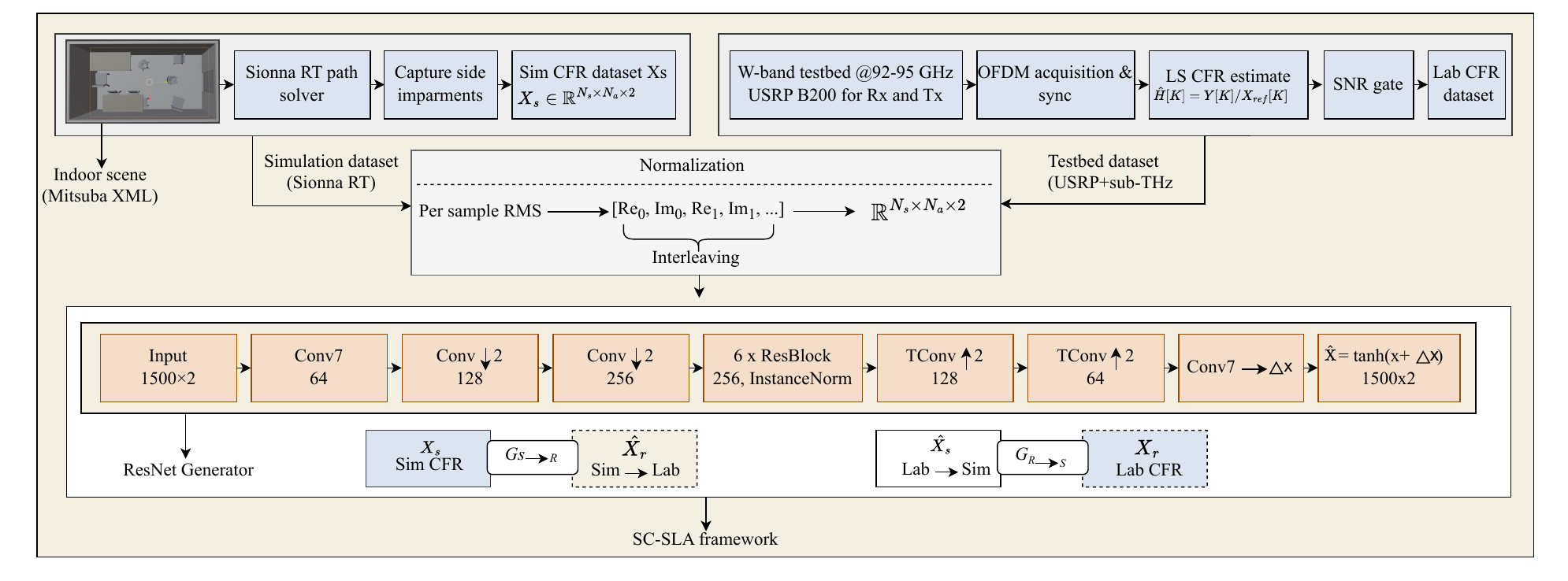}
\caption{SC-SLA framework. The simulation branch (top left) runs the
Sionna~RT path solver on an indoor Mitsuba scene, adds capture-side
impairments, and yields the CT dataset $X_s\in\mathbb{R}^{N_s\times N_{\text{a}}\times 2}$. The testbed branch (top right) feeds the W-band USRP-B200 chain through OFDM synchronization, LS CFR estimation $\hat H[k]=Y[k]/X_{\text{ref}}[k]$, and an SNR gate, producing the measured dataset. Both datasets share a per-sample RMS normalization with Re/Im
interleaving (middle). The 1D ResNet generator (bottom) is an
encoder--residual-trunk--decoder with a $\tanh$-bounded residual delta head
$\hat{\mathbf{x}}=\tanh(\mathbf{x}+\Delta\mathbf{x})$ for an input CFR sample $\mathbf{x}$, forming the forward map
$G_{S\to R}\!:X_s\!\to\!\hat X_r$ and the reverse map
$G_{R\to S}\!:X_r\!\to\!\hat X_s$ in the cycle-consistent translator.}
\label{fig:overview}
\end{figure*}
SC-SLA calibrates the ray-traced CT dataset $X_s$ toward the measured dataset $X_r$. The proposed SC-SLA adopts a two-generator, cycle-consistent translation structure inspired by CycleGAN~\cite{cyclegan}, but removes the adversarial discriminators and replaces them with batch-level channel-statistics losses. The result is a GAN-inspired, non-adversarial, unpaired channel translator. For a normalized simulated CFR sample $\mathbf{x}_s\in\mathbb{R}^{N_{\text{a}}\times 2}$, the forward generator
$G_{S\to R}:\mathbb{R}^{N_{\text{a}}\times2}\to\mathbb{R}^{N_{\text{a}}\times 2}$ maps the simulated sample to a real-world-like sample, $\hat{\mathbf{x}}_r=G_{S\to R}(\mathbf{x}_s)$. Applying this map to the full simulated dataset gives the calibrated dataset $\hat X_r=G_{S\to R}(X_s)$. For a normalized measured CFR sample $\mathbf{x}_r\in\mathbb{R}^{N_{\text{a}}\times 2}$, a reverse generator $G_{R\to S}$ maps measured samples back to the ray-traced domain, $\hat{\mathbf{x}}_s=G_{R\to S}(\mathbf{x}_r)$. Applying the reverse map to all measured samples gives $\hat X_s=G_{R\to S}(X_r)$, which is used for cycle-consistency regularization. Both generators share the same architecture.

The calibration objective is distributional. SC-SLA does not force a simulated CFR to match a particular measured CFR, because the datasets are not synchronized and no sample-wise correspondence exists. Instead, the forward map is constrained to reproduce the channel statistics that govern OFDM receiver design, including the delay-domain energy distribution, $\tau_{\text{rms}}$, and frequency-domain magnitude response. This matches the data collection process, where simulated and measured datasets represent the same nominal static link and OFDM grid. However, their sample-to-sample variations arise independently. Fig.~\ref{fig:overview} depicts the data flow and the complete SC-SLA framework.

In an adversarial translator, discriminators learn implicit criteria for separating translated outputs from samples in the corresponding target domains, and the generators are updated through a minimax objective. SC-SLA instead minimizes explicit, differentiable discrepancies in the PDP, $\tau_{\text{rms}}$ distribution, and CFR-magnitude profile because these receiver-relevant target statistics are available from the measured captures. Our non-adversarial formulation removes discriminator design and minimax tuning while making the physical role of each alignment term explicit. The reverse generator $G_{R\to S}$ does not replace a discriminator. We retain $G_{R\to S}$ to provide the inverse path required by cycle-consistency regularization. We use only $G_{S\to R}$ to calibrate simulated CFRs during inference.

\subsection{Generator Architecture}
\label{sec:generator}

The SC-SLA generator is a 1D ResNet~\cite{he2016} with an encoder--residual-trunk--decoder structure, illustrated in the bottom panel of Fig.~\ref{fig:overview}. The topology adapts the translation generator of Johnson \emph{et al.}~\cite{johnson2016}, also used by CycleGAN~\cite{cyclegan}, from 2D images to the 1D active CFR. The real and imaginary components of each normalized sample form two input channels over the $N_{\text{a}}=1500$ active subcarriers. Since SC-SLA retains the CycleGAN translation structure and removes only its adversarial discriminators, the generator family is kept fixed and the layer dimensions are set by the OFDM numerology in Section~\ref{sec:signal}. A $7$-tap input convolution lifts the two-channel input to $64$ feature channels over a spectral aperture of $7\Delta f\approx\SI{171}{\kHz}$. By Eq.~\eqref{eq:cfr}, a path at excess delay $\tau$ varies across frequency with period $1/\tau$. Consequently, the longest delay retained in Eq.~\eqref{eq:cir}, \SI{400}{\nano\second}, induces the fastest CFR ripple with a \SI{2.5}{\MHz} period. The input aperture covers at most $7\%$ of this cycle, so it extracts local CFR level, slope, and curvature while leaving multipath-induced ripples to deeper layers. Two stride-2 convolutions with $3$-tap kernels then reduce the active-subcarrier length from $1500$ to $750$ and $375$, while increasing the channel dimension from $64$ to $128$ and $256$. Two strided stages provide the deepest exact compression of the active grid. A third halving would require a fractional length of $187.5$. The channel doubling keeps the activation volume constant ($1500\times64=750\times128=375\times256=96{,}000$ values per CFR), trading spectral resolution for feature richness without reducing representational capacity~\cite{he2016}.

After the strided stages, the compressed spectral grid contains one feature vector per four subcarriers, or approximately \SI{97.7}{\kHz} of spectrum. A residual trunk of six instance-normalized blocks refines the encoded CFR features at this resolution, matching the CycleGAN depth for inputs of comparable size~\cite{cyclegan}. Across the input convolution, strided stages, twelve trunk convolutions, mirrored decoder, and output head, the end-to-end receptive field reaches $127$ subcarriers, or approximately \SI{3.1}{\MHz}. This span exceeds the \SI{2.5}{\MHz} period of the fastest ripple, so each output subcarrier observes at least one complete cycle of the finest spectral structure generated by the truncated delay window. Without downsampling, the same depth would cover only $45$ subcarriers, or approximately \SI{1.1}{\MHz}. The receptive field is also close to the $150$-subcarrier slices used by the sub-band PDP loss in Section~\ref{sec:losses}, so learned corrections act at the spectral granularity penalized by the loss. Instance normalization~\cite{ulyanov2016} standardizes each capture individually rather than across the mini-batch, which is appropriate for unpaired translation with capture-dependent receive power.

The decoder mirrors the encoder and restores the original $1500$-subcarrier length through two stride-2 transposed convolutions that reduce the channel dimension from $256$ back to $64$, with explicit length matching for the odd $375\rightarrow750$ upsampling. A final un-normalized $7$-tap convolutional head predicts the residual correction $\Delta\mathbf{x}$ on the amplitude scale of the input CFR. For a generic normalized input sample $\mathbf{x}\in\mathbb{R}^{N_{\text{a}}\times2}$ from either domain, the output is formed as $\hat{\mathbf{x}}=\tanh(\mathbf{x}+\Delta\mathbf{x})$. The residual skip biases the map toward the identity, so the network refines the ray-traced CFR rather than reconstructing it from scratch. The $\tanh$ head confines the output to the normalized $[-1,1]$ support shared by both datasets. Since the generator is fully feed-forward, each calibrated CFR requires only one forward pass, allowing SC-SLA to be inserted into a W-band link-level simulator without iterative calibration.

\subsection{Channel-Statistics Losses}
\label{sec:losses}

The generator architecture determines the form of the CFR correction, while the loss function determines which channel properties it preserves or reproduces. We use two types of constraints. The first group consists of channel-statistics losses, which compare ensemble-level quantities computed over independently sampled simulated and measured mini-batches. These losses are not applied sample by sample, because the captures are unpaired and no measured realization corresponds to a particular simulated one. The quantities most relevant to OFDM receiver design, including PDP shape, $\tau_{\text{rms}}$, and the average subcarrier-magnitude profile, are distributional channel properties rather than deterministic labels for individual samples.

The second group consists of instantaneous regularizers, namely the cycle and identity losses. These losses act directly on each sample and prevent the learned mapping from becoming an unconstrained distribution-matching transform. A purely statistical objective is insensitive to permutations within a mini-batch and cannot alone ensure that each translated CFR remains tied to the geometry and structure of its input. The cycle and identity losses therefore complement the residual generator parameterization in Section~\ref{sec:generator}. We retain multiple statistical losses because each constrains a different receiver-relevant projection of the channel. The ablation results in Section~\ref{sec:ablation} demonstrate that no single term subsumes the others.

Let $M$ denote the mini-batch size and $T=20$ the number of retained CIR taps within the common \SI{400}{\nano\second} delay window. We denote a simulated mini-batch by
$\mathcal{B}_s \in \mathbb{R}^{M \times N_{\text{a}} \times 2}$ and an independently drawn measured mini-batch of the same size by $\mathcal{B}_r$. For a CFR mini-batch $\mathcal{B}$, let $P(\mathcal{B}) \in \mathbb{R}^{M \times T}$ be the collection of per-sample normalized PDPs obtained using the CIR and PDP definitions in Section~\ref{sec:sysmodel}. The losses below compare mini-batch statistics rather than enforcing arbitrary one-to-one matching between simulated and measured samples.

\noindent\textbf{Truncated PDP loss:}
The first term matches the average truncated PDP of the measured batch to that of the translated simulated batch as
\begin{equation}
\mathcal{L}_{\text{PDP}} =
\frac{1}{T}
\big\| \, \overline{P(\mathcal{B}_r)}
- \overline{P(G_{S\to R}(\mathcal{B}_s))} \, \big\|_1,
\label{eq:lpdp}
\end{equation}
where $\overline{(\cdot)}$ denotes averaging over the mini-batch.

\noindent\textbf{Sub-band PDP loss:}
To constrain local frequency-dependent delay behavior, the $N_{\text{a}}$ active subcarriers are divided into $K=10$ contiguous slices, each containing $N_{\text{a}}/K=150$ subcarriers, or approximately $\SI{3.66}{\MHz}$. For each slice $c$, we compute a 150-point IFFT and form a sub-band PDP $P_c(\cdot)$ from the first $T$ taps. Since this sub-band transform has a tap spacing of
$1/(\SI{3.66}{\MHz}) \approx \SI{273}{\nano\second}$, the sub-band transform provides a coarser delay grid than the global $\SI{20}{\nano\second}$ CIR. The resulting loss captures slower frequency-selective structure over a wider effective delay window and is given by
\begin{equation}
\mathcal{L}_{\text{sPDP}} =
\frac{1}{KT}
\sum_{c=1}^{K}
\big\| \, \overline{P_c(\mathcal{B}_r)}
- \overline{P_c(G_{S\to R}(\mathcal{B}_s))} \, \big\|_1.
\end{equation}
The factor $1/(KT)$ averages the absolute PDP discrepancy over the $K$ sub-bands and $T$ retained taps.

\noindent\textbf{$\tau_{\text{rms}}$ moment loss:}
Let $\mu_{\tau}(\cdot)$ and $\sigma_{\tau}(\cdot)$ denote the mini-batch mean and standard deviation of the per-sample $\tau_{\text{rms}}$ values defined in Eq.~\eqref{eq:taurms}. Since these are scalar batch statistics, we penalize their mismatch using absolute differences as
\begin{align}
\mathcal{L}_{\tau} &=
\big| \mu_{\tau}(\mathcal{B}_r)
- \mu_{\tau}(G_{S\to R}(\mathcal{B}_s)) \big| \nonumber\\
&\quad
+ 2 \big| \sigma_{\tau}(\mathcal{B}_r)
- \sigma_{\tau}(G_{S\to R}(\mathcal{B}_s)) \big|.
\end{align}
Here, the standard-deviation term receives twice the weight of the mean term to emphasize the dispersion error identified in Section~\ref{sec:intro}, where the simulated twin exhibits a narrower capture-to-capture $\tau_{\text{rms}}$ distribution than the measured link.

\noindent\textbf{$\tau_{\text{rms}}$ quantile loss:}
The moment loss controls only the first two summary statistics. To better align the full empirical distribution, especially the upper tail, we also match the sorted $\tau_{\text{rms}}$ values. Let $\tilde{\boldsymbol{\tau}}(\mathcal{B})\in\mathbb{R}^{M}$ denote the vector of per-sample $\tau_{\text{rms}}$ values from $\mathcal{B}$ in ascending order. We define
\begin{equation}
\mathcal{L}_{\tau q} =
\frac{1}{M}
\big\| \tilde{\boldsymbol{\tau}}(\mathcal{B}_r)
- \tilde{\boldsymbol{\tau}}(G_{S\to R}(\mathcal{B}_s)) \big\|_1.
\end{equation}
For equal-sized, equally weighted mini-batches, $\mathcal{L}_{\tau q}$ equals the empirical 1-Wasserstein distance between the measured and generated $\tau_{\text{rms}}$ samples in Eq.~\eqref{eq:wasserstein1}. Fig.~\ref{fig:cir_statistics_gap} reports this distance for the held-out pre-calibration data, whereas training evaluates it on independently drawn mini-batches.

\noindent\textbf{CFR-magnitude NMSE loss:}
The delay-domain terms constrain the PDP and $\tau_{\text{rms}}$ statistics. However, they do not fully determine the average spectral profile across active subcarriers. We therefore include a frequency-domain magnitude loss. For a mini-batch $\mathcal{B}$ represented by real and imaginary channels, let $|\mathcal{B}|$ denote the corresponding per-sample, per-subcarrier magnitude. The loss is
\begin{equation}
\mathcal{L}_{\text{mag}} =
\frac{
\big\| \overline{|\mathcal{B}_r|}
- \overline{|G_{S\to R}(\mathcal{B}_s)|} \big\|_2^2
}{
\big\| \overline{|\mathcal{B}_r|} \big\|_2^2
}.
\end{equation}

\noindent\textbf{Cycle and identity losses:}
Cycle and identity regularization keep the unpaired translator close to a physically meaningful correction rather than an unconstrained remapping. We define
\begin{align}
\mathcal{L}_{\text{cyc}} &=
\|G_{R\to S}(G_{S\to R}(\mathcal{B}_s)) - \mathcal{B}_s\|_1 \nonumber\\
&\quad +
\|G_{S\to R}(G_{R\to S}(\mathcal{B}_r)) - \mathcal{B}_r\|_1, \\
\mathcal{L}_{\text{id}} &=
\|G_{S\to R}(\mathcal{B}_r) - \mathcal{B}_r\|_1
+
\|G_{R\to S}(\mathcal{B}_s) - \mathcal{B}_s\|_1.
\end{align}
The cycle-consistency loss requires a sample translated to the other domain and mapped back to reconstruct the original input. Cycle consistency therefore preserves sample-specific CFR structure and discourages many-to-one mappings. However, this loss constrains the compositions $G_{R\to S}\circ G_{S\to R}$ and $G_{S\to R}\circ G_{R\to S}$, while the individual generators may still learn compensating transformations that cancel under composition. The identity loss reduces this ambiguity by regularizing each generator toward the identity mapping for inputs that already belong to its output domain. Identity regularization limits unnecessary amplitude and phase changes to CFRs that already exhibit the intended domain statistics. In the present static single-link setting, the bounded residual head provides much of the same input--output coupling even when the cycle and identity losses are removed, as quantified in Section~\ref{sec:ablation}. We nevertheless retain both losses as conservative sample-level regularizers.

The complete generator objective is
\begin{align}
\mathcal{L}_G &=
\lambda_{\text{cyc}} \mathcal{L}_{\text{cyc}}
+ \lambda_{\text{id}} \mathcal{L}_{\text{id}}
+ \lambda_{\tau} \mathcal{L}_{\tau}
+ \lambda_{\tau q} \mathcal{L}_{\tau q} \nonumber \\
&\quad
+ \lambda_{\text{PDP}} \mathcal{L}_{\text{PDP}}
+ \lambda_{\text{sPDP}} \mathcal{L}_{\text{sPDP}}
+ \lambda_{\text{mag}} \mathcal{L}_{\text{mag}},
\label{eq:lg}
\end{align}
where $\boldsymbol{\lambda}$ is the vector of scalar loss weights listed in Table~\ref{tab:hyperparams}. All SC-SLA loss weights are selected once using the 95~GHz training/validation split and kept fixed for the in-domain test, cross-frequency transfer, and ablation experiments. The weighting follows a coarse validation calibration in which the cycle and identity terms provide conservative sample-level regularization, while the delay-domain terms receive larger weights to address the PDP and $\tau_{\text{rms}}$ mismatches discussed in Section~\ref{sec:intro}.

\subsection{Training Procedure}
\label{sec:training}

We implement the mappings as separate ResNet generators with independent parameter sets $\theta_{S\to R}$ and $\theta_{R\to S}$. Consequently, we learn $G_{R\to S}$ directly through the overall objective rather than obtain it by inverting $G_{S\to R}$. During each mini-batch, we evaluate both networks, and a common Adam optimizer~\cite{kingma2015} updates both parameter sets by minimizing $\mathcal{L}_G$. The reverse generator receives gradients through the cycle and identity terms, while the forward generator is additionally constrained by the channel-statistics losses. Training is therefore bidirectional, whereas sim-to-lab inference uses only $G_{S\to R}$.

At 95~GHz, $45{,}000$ simulated and measured captures are used for optimization, and $5{,}000$ captures are held-out from training. Both generators are trained for $N_{\mathrm{ep}}=300$ epochs with a learning rate of $10^{-4}$, linear decay after $N_{\mathrm{dec}}=150$ epochs, BF16 automatic mixed precision, and an $\ell_2$ gradient-norm clip of $1.0$. Checkpoint selection, final evaluation, and the inference benchmark use FP32. Table~\ref{tab:hyperparams} lists the system, RT, training, and model-selection parameters.

\begin{table}[t]
\renewcommand{\arraystretch}{1.1}
\setlength{\tabcolsep}{4.5pt}
\caption{System, RT, training, and model-selection parameters.}
\label{tab:hyperparams}
\centering
\begin{tabular}{@{}>{\raggedright\arraybackslash}p{0.49\columnwidth}>{\raggedright\arraybackslash}p{0.47\columnwidth}@{}}
\toprule
Parameter & Value \\
\midrule
\multicolumn{2}{c}{\itshape System Parameters} \\
\midrule
Carrier frequencies & $92$--$95$~GHz \\
IF / LO frequency (GHz) & $\{2,3,4,5\}$ / $11.25$ ($\times 8$) \\
FFT size / active subcarriers & $2048$ / $1500$ \\
Cyclic-prefix length & $512$ samples \\
Sampling bandwidth / $\Delta f$ & $50$~MHz / $24.41$~kHz \\
Occupied bandwidth & $36.6$~MHz \\
CIR taps & $T=20$ at $20$~ns ($400$~ns window) \\
TX--RX separation & $1.75$~m (LOS) \\
Antennas & WR-10 pyramidal horns, co-polarized \\
Software-defined radio & USRP-B200, $50$~MHz master clock \\
Receiver gain / SNR gate & $70$~dB / $8$~dB \\
Capture burst per carrier & $15$~s \\
Captures per carrier ($N_s$, $N_r$) & $50{,}000$ each \\
Training frequency & $95$~GHz\\
Testing frequency & $92$--$95$~GHz\\ 
\midrule
\multicolumn{2}{c}{\itshape Ray-Traced CT Parameters} \\
\midrule
Path solver & Sionna~RT \\
Enabled interactions & LOS, specular, refraction \\
Max interaction depth & $5$ (diffuse reflection disabled) \\
Antenna power pattern & Cosine-power, $n_{\mathrm{h}}=126.18$, $12^{\circ}$ 3-dB beamwidth \\
Timing jitter (std) & $2.5$~ns \\
Common phase jitter (std) & $0.20$~rad \\
Per-capture SNR & $28$~dB mean, $2$~dB std \\
\midrule
\multicolumn{2}{c}{\itshape SC-SLA Hyperparameters} \\
\midrule
Clipping quantile $\rho_{\text{clip}}$ & $0.995$ \\
Sub-band count $K$ & $10$ \\
Residual blocks / channel widths & $6$ / $64$--$128$--$256$ \\
Optimizer & Adam ($\beta_1=0.5$, $\beta_2=0.999$) \\
Learning rate / decay start $N_{\mathrm{dec}}$ & $10^{-4}$ / epoch $150$ \\
Training epochs $N_{\mathrm{ep}}$ / batch size $M$ & $300$ / $64$ \\
Precision / gradient-norm clip & BF16 / $1.0$ \\
$\lambda_{\text{cyc}},\ \lambda_{\text{id}}$ & $10,\ 2$ \\
$\lambda_{\tau},\ \lambda_{\tau q}$ & $5,\ 3$ \\
$\lambda_{\text{PDP}},\ \lambda_{\text{sPDP}},\ \lambda_{\text{mag}}$ & $15,\ 20,\ 1$ \\
95~GHz training samples & $45{,}000$\\
Reserved samples & $5{,}000$\\
Model-selection interval $N_{\mathrm{sel}}$ & every $5$ epochs \\
$J$ coefficients $\omega_{\mathrm{PDP}},\ \omega_{\mathrm{mag}}$ & $50,\ 0.25$ \\
Evaluation subsample & $2{,}000$ captures per carrier \\
\bottomrule
\end{tabular}
\end{table}
\begin{algorithm}[t]
  \caption{SC-SLA training}
  \label{alg:pcsla}
  \begin{algorithmic}[1]
  \State \textbf{Input:} $X_s,X_r,\boldsymbol{\lambda},N_{\mathrm{ep}},N_{\mathrm{sel}},\rho_{\text{clip}}$
  \State Split $X_s$ and $X_r$ into optimization and reserved subsets
  \State Initialize $G_{S\to R}$ and $G_{R\to S}$ with parameters $\theta_{S\to R}$ and $\theta_{R\to S}$
  \State Set $q_{\text{clip}}$ to the $\rho_{\text{clip}}$-quantile of the pooled 95~GHz magnitudes
  \State Set $J^{\star}\leftarrow\infty$
  \For{$e=1,\dots,N_{\mathrm{ep}}$}
    \State Form independently shuffled mini-batch index sets $\{\pi_s^{(i)}\}$ and $\{\pi_r^{(i)}\}$
    \For{each mini-batch $i$}
      \State $\mathcal{B}_s\!\leftarrow\!X_s[\pi_s^{(i)}]$, $\mathcal{B}_r\!\leftarrow\!X_r[\pi_r^{(i)}]$
      \State Evaluate both mappings and compute $\mathcal{L}_G$ from Eqs.~\eqref{eq:lpdp}--\eqref{eq:lg}
      \State Compute $\nabla_{\theta_{S\to R},\theta_{R\to S}}\mathcal{L}_G$
      \State Clip $\|\nabla\mathcal{L}_G\|_2\!\le\!1$ and jointly update both generators with Adam
    \EndFor
    \If{$e$ is divisible by $N_{\mathrm{sel}}$}
      \State Evaluate $D_{\mathrm{KS}}$, $\operatorname{NMSE}_{\mathrm{PDP}}$, and $\operatorname{NMSE}_{\mathrm{mag}}$ on reserved captures
      \State Compute $J=D_{\mathrm{KS}}+\omega_{\mathrm{PDP}}\,\operatorname{NMSE}_{\mathrm{PDP}}$
      \Statex \hspace{\algorithmicindent}$+\omega_{\mathrm{mag}}\,\operatorname{NMSE}_{\mathrm{mag}}$
      \If{$J<J^{\star}$}
        \State $J^{\star}\leftarrow J$ and save the current $G_{S\to R}$ checkpoint
      \EndIf
    \EndIf
  \EndFor
  \State \textbf{return} $G_{S\to R}$ checkpoint with minimum $J$
  \end{algorithmic}
  \end{algorithm}

Checkpoint selection uses three lower-is-better metrics evaluated on reserved captures. Let $\mathcal{E}_g$ denote the output CFR set of a candidate model and $\mathcal{E}_r$ the measured reference set. Both are evaluation sets rather than the training mini-batches $\mathcal{B}_s$ and $\mathcal{B}_r$ of Section~\ref{sec:losses}. Let $\widehat{F}_g(t)$ and $\widehat{F}_r(t)$ denote the empirical cumulative distribution functions (CDFs) of the per-sample $\tau_{\text{rms}}$ values for $\mathcal{E}_g$ and $\mathcal{E}_r$, respectively. Their two-sample KS statistic is
\begin{equation}
D_{\mathrm{KS}}
=
\sup_{t\ge 0}
\left|
\widehat{F}_g(t)-\widehat{F}_r(t)
\right|.
\label{eq:ks}
\end{equation}
Here $\sup$ denotes the supremum, that is, the least upper bound of $|\widehat{F}_g(t)-\widehat{F}_r(t)|$ over all delay-spread values $t\ge 0$. Both empirical CDFs are right-continuous step functions with finitely many jumps. The difference therefore takes finitely many values, and the supremum is attained as a maximum over the pooled $\tau_{\text{rms}}$ values of $\mathcal{E}_g$ and $\mathcal{E}_r$. Accordingly, $D_{\mathrm{KS}}$ is the largest vertical gap between the two empirical CDFs and is bounded to $[0,1]$. The two NMSE metrics are
\begin{align}
\operatorname{NMSE}_{\mathrm{PDP}}
&=
\frac{
\big\|\overline{P(\mathcal{E}_g)}-\overline{P(\mathcal{E}_r)}\big\|_2^2
}{
\big\|\overline{P(\mathcal{E}_r)}\big\|_2^2
},
\label{eq:nmse_pdp}\\
\operatorname{NMSE}_{\mathrm{mag}}
&=
\frac{
\big\|\overline{|\mathcal{E}_g|}-\overline{|\mathcal{E}_r|}\big\|_2^2
}{
\big\|\overline{|\mathcal{E}_r|}\big\|_2^2
}.
\label{eq:nmse_mag}
\end{align}
The metric $\operatorname{NMSE}_{\mathrm{mag}}$ has the same mathematical form as the training loss $\mathcal{L}_{\text{mag}}$, but it is evaluated on the reserved or held-out sets $\mathcal{E}_g$ and $\mathcal{E}_r$ rather than on training mini-batches. In contrast, $\operatorname{NMSE}_{\mathrm{PDP}}$ uses normalized squared $\ell_2$ error between ensemble-mean PDPs, whereas $\mathcal{L}_{\text{PDP}}$ uses mean absolute error. The PDP metric therefore provides an in-objective consistency check without duplicating the training loss.

The three metrics form the scalar model-selection criterion
\begin{equation}
J
=
D_{\mathrm{KS}}
+\omega_{\mathrm{PDP}}\,\operatorname{NMSE}_{\mathrm{PDP}}
+\omega_{\mathrm{mag}}\,\operatorname{NMSE}_{\mathrm{mag}}.
\label{eq:model_selection}
\end{equation}
The fixed coefficients $\omega_{\mathrm{PDP}}=50$ and $\omega_{\mathrm{mag}}=0.25$ weight the two NMSE terms relative to $D_{\mathrm{KS}}$ during checkpoint selection. They act only on the selection criterion and are distinct from the training-loss weights $\boldsymbol{\lambda}$ in Eq.~\eqref{eq:lg}. The coefficient $\omega_{\mathrm{PDP}}$ compensates for the small numerical range of $\operatorname{NMSE}_{\mathrm{PDP}}$, whereas $\omega_{\mathrm{mag}}$ limits the contribution of the numerically larger $\operatorname{NMSE}_{\mathrm{mag}}$. Both values are fixed once on the 95~GHz validation split and reused for every reported experiment. The statistic $D_{\mathrm{KS}}$ enters $J$ as a sup-norm on the $\tau_{\text{rms}}$ CDF. No training term takes this form, so it scores the distribution targeted by $\mathcal{L}_{\tau}$ and $\mathcal{L}_{\tau q}$ under a different functional and penalizes checkpoints that match the moments and quantiles while leaving a localized gap in the CDF.

Every $N_{\mathrm{sel}}=5$ epochs, the checkpoint is evaluated on a fixed subset of reserved captures using the criterion $J$. The checkpoint with the lowest $J$ is retained. The reported 95~GHz results are therefore in-domain, model-selected results on non-training captures. The 92--94~GHz carriers are used only for cross-frequency evaluation, not for optimization, model selection, or carrier-specific tuning. Algorithm~\ref{alg:pcsla} summarizes the joint generator optimization and checkpoint selection.

We conduct all simulations on an Ubuntu 24.04 LTS workstation equipped with an AMD Ryzen 9 9950X (16 cores, 32 threads, 5.7 GHz) processor, 128 GB RAM, and an NVIDIA RTX 4000 Ada Generation GPU with 20 GB memory. We implement SC-SLA in Python 3.12 with PyTorch 2.8.

\section{Experiments and Result Analysis}
\label{sec:experiments}

\subsection{Baselines and Evaluation Metrics}
\label{sec:baselines}

We compare SC-SLA with four supervised neural generators that use the same normalization method and the same zero-initialized residual-delta form, $\hat{\mathbf{x}}=\mathbf{x}+\Delta\mathbf{x}$. The baselines include a fully connected neural network (FCNN) with four $512$-unit layers, LayerNorm, and dropout. The one-dimensional convolutional neural network (CNN1D) has six residual convolutional blocks and $128$ filters. The three-layer bidirectional long short-term memory network (BiLSTM) has $128$ hidden units. The depth-$2$ one-dimensional U-Net (UNet1D)~\cite{ronneberger2015} uses $64$ base channels.

The supervised baselines are constrained by the data collection process. Simulated and measured captures are triggered independently, so no physical one-to-one correspondence exists between a simulated CFR and a measured CFR. We therefore train the baselines using Huber regression~\cite{huber1964robust} on index-aligned pairs, together with weak batch regularizers on CFR magnitude, sub-band magnitude, and $\tau_{\text{rms}}$ statistics. We use this Huber-based configuration as the strongest paired-supervision setting supported by the data and as a direct comparison with the unpaired SC-SLA objective. All models are selected using the same composite validation criterion, with family-specific weights chosen to place the constituent terms on comparable numerical scales. The three lower-is-better metrics are defined in Section~\ref{sec:training}. They comprise $D_{\mathrm{KS}}$, $\operatorname{NMSE}_{\mathrm{PDP}}$, and $\operatorname{NMSE}_{\mathrm{mag}}$. The KS statistic is not directly optimized by any training loss. The PDP metric compares the same ensemble-mean PDPs as $\mathcal{L}_{\text{PDP}}$ but uses normalized squared $\ell_2$ error and therefore serves mainly as an in-objective consistency check. The CFR-magnitude metric measures preservation of the average frequency-domain magnitude profile.
\begin{figure}[t]
\centering
\includegraphics[width=\columnwidth]{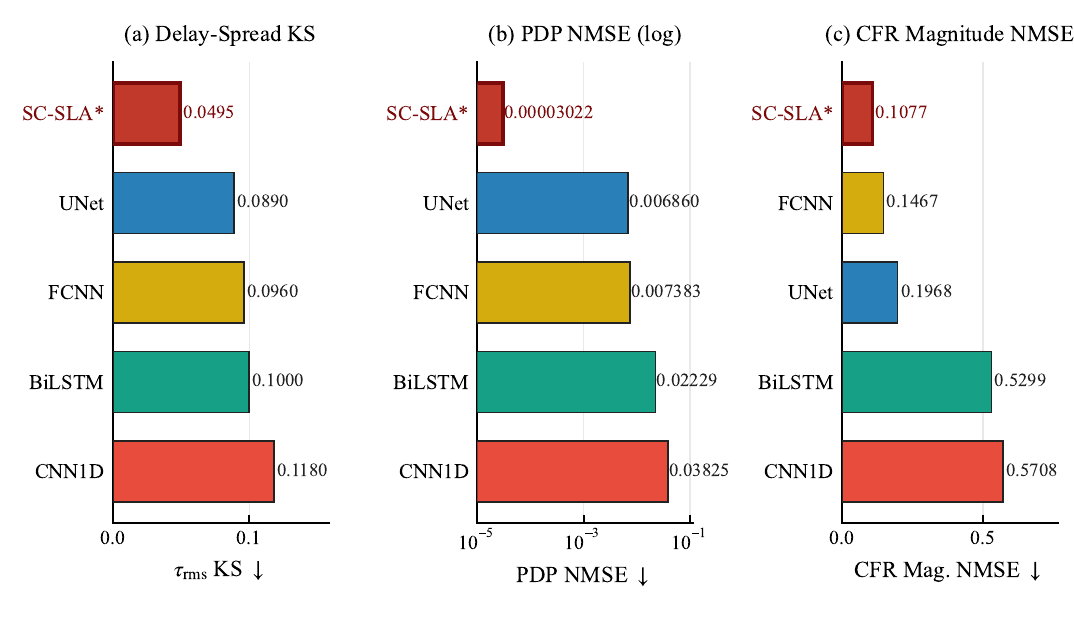}
\caption{In-domain calibration at 95~GHz on $2{,}000$ held-out captures. The panels report (a) the $\tau_{\text{rms}}$ KS statistic, (b) truncated PDP NMSE on a logarithmic axis, and (c) per-subcarrier CFR-magnitude NMSE across SC-SLA and the four supervised baselines. SC-SLA achieves the lowest error on all three metrics, with KS $0.0495$ and PDP NMSE $3.0\times10^{-5}$.}
\label{fig:indomain}
\end{figure}
\begin{figure}[t]
\centering
\includegraphics[width=0.95\columnwidth]{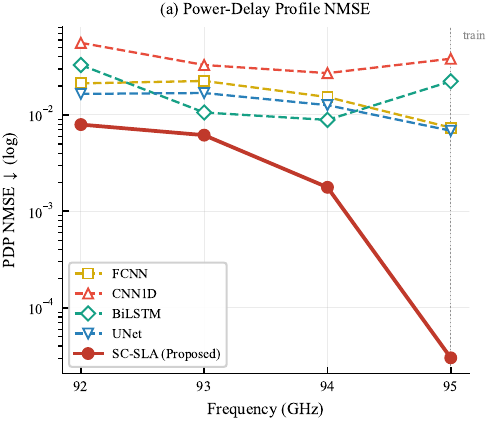}
\caption{Cross-frequency truncated PDP NMSE at 92--94~GHz using the 95~GHz checkpoint without retraining. The 95~GHz point is the in-domain carrier. SC-SLA achieves the lowest PDP error at every held-out carrier, with $2.1\times$, $1.7\times$, and $5.0\times$ gains over the strongest supervised baseline at 92, 93, and 94~GHz, respectively.}
\label{fig:crossfreq_pdp}
\end{figure}

\begin{figure}[!t]
\centering
\includegraphics[width=0.95\columnwidth]{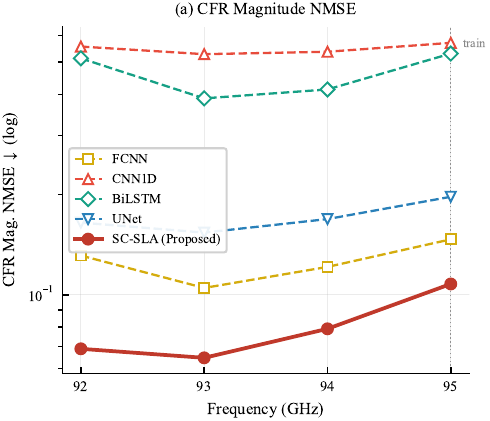}
\caption{Cross-frequency per-subcarrier CFR-magnitude NMSE at 92--94~GHz using the 95~GHz checkpoint without retraining. The 95~GHz point is the in-domain carrier. SC-SLA preserves the magnitude profile under carrier shift and remains below every supervised baseline at each held-out carrier.}
\label{fig:crossfreq_mag}
\end{figure}
\subsection{In-Domain Results at 95~GHz}
\label{sec:indomain}

\begin{figure*}[t]
\centering
\includegraphics[width=\textwidth]{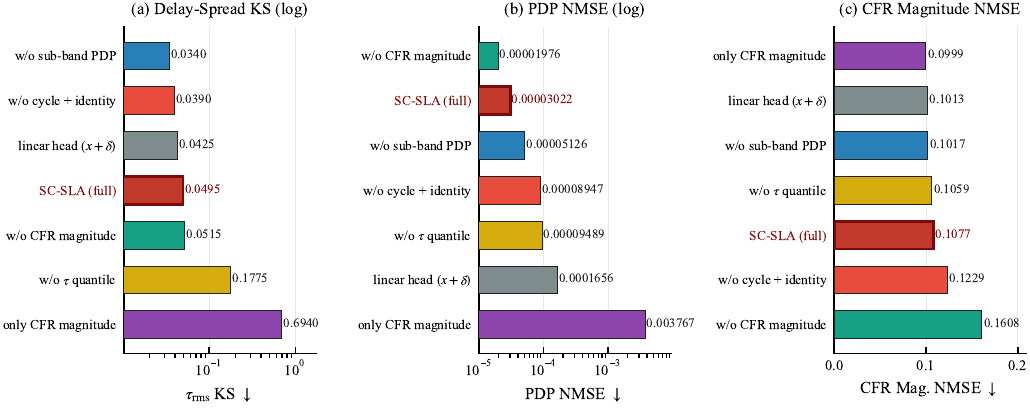}
\caption{In-domain ablation at 95~GHz. The panels report (a) $\tau_{\text{rms}}$ KS statistic, (b) truncated PDP NMSE, and (c) per-subcarrier CFR-magnitude NMSE across the full model and six single-component variants. Bars are sorted within each panel. Panels (a) and (b) use logarithmic axes. No ablated variant dominates all three metrics, while the full model avoids severe degradation in any metric.}
\label{fig:abl_indomain}
\end{figure*}
\begin{figure*}[t]
\centering
\includegraphics[width=0.95\textwidth]{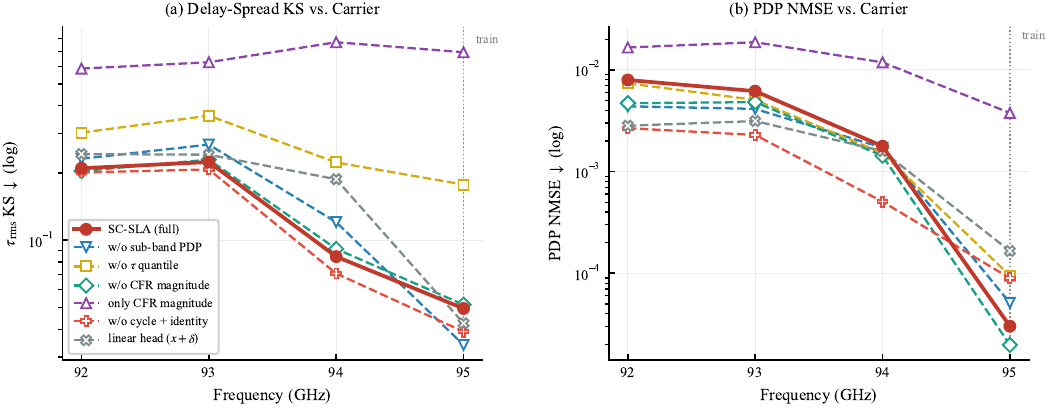}
\caption{Cross-frequency ablation using the 95~GHz checkpoint without retraining. The 95~GHz point is the in-domain held-out split. Panel (a) plots the $\tau_{\text{rms}}$ KS statistic and panel (b) plots truncated PDP NMSE versus carrier, both on logarithmic axes. Removing the $\tau_{\text{rms}}$ quantile loss or all delay-domain terms consistently worsens the $\tau_{\text{rms}}$ match, with the clearest separation at 94~GHz.}
\label{fig:abl_crossfreq}
\end{figure*}


We first evaluate SC-SLA and the four supervised baselines on $2{,}000$ held-out captures at $95$~GHz. For context, the uncalibrated twin, denoted as ``Sionna (identity)'', applies the impairment-augmented ray-traced output directly through $G_{\mathrm{id}}(\mathbf{x})=\mathbf{x}$ for an input CFR sample $\mathbf{x}$. Fig.~\ref{fig:cir_statistics_gap} shows that this input has a large $\tau_{\text{rms}}$-distribution mismatch, with $D_{\mathrm{KS}}=0.861$. Thus, the impairment-augmented twin alone does not reproduce the measured delay-domain statistics.

The supervised baselines reduce the $\tau_{\text{rms}}$ KS statistic to $0.0890$--$0.1180$, as shown in Fig.~\ref{fig:indomain}. However, CNN1D and BiLSTM yield CFR-magnitude NMSE values of $0.5708$ and $0.5299$, respectively. This spectral distortion is consistent with the limitation of index-aligned supervision because the independently triggered simulated and measured captures are not physically paired.
SC-SLA achieves the lowest error across all three metrics in Fig.~\ref{fig:indomain}. It reduces the KS statistic to $0.0495$, a $44\%$ reduction from the $0.0890$ obtained by UNet1D, the best supervised baseline on this metric. Furthermore, SC-SLA attains a CFR-magnitude NMSE of $0.1077$ and a PDP NMSE of $3.0\times10^{-5}$. The latter is more than two orders of magnitude below every supervised baseline. Because PDP NMSE compares the same ensemble-mean PDPs as $\mathcal{L}_{\text{PDP}}$ using a different error norm, it serves as an in-objective consistency check rather than a fully independent metric.

\subsection{Cross-Frequency Generalization at 92--94~GHz}
\label{sec:crossfreq}

We next examine whether the calibration learned at $95$~GHz transfers to nearby W-band carriers. For this experiment, we apply the same $95$~GHz checkpoint directly to Sionna inputs at $92$, $93$, and $94$~GHz, without retraining, fine-tuning, or carrier-specific model selection. We then compare the calibrated outputs with $2{,}000$ measured captures at each carrier.

The PDP NMSE increases when the model is evaluated away from the training carrier, rising from the in-domain value of $3.0\times10^{-5}$ to the $10^{-3}$ range. This degradation is expected because no $92$--$94$~GHz data are used during training. Even under this shift, SC-SLA achieves the lowest PDP NMSE at every held-out carrier. Compared with the strongest supervised baseline at each frequency, SC-SLA improves the PDP NMSE by $2.1\times$, $1.7\times$, and $5.0\times$ at $92$, $93$, and $94$~GHz, respectively (Fig.~\ref{fig:crossfreq_pdp}). SC-SLA also achieves the lowest CFR-magnitude NMSE across all three held-out carriers (Fig.~\ref{fig:crossfreq_mag}).

The $\tau_{\text{rms}}$ results follow the same overall trend, although the margin varies by carrier. SC-SLA keeps the $\tau_{\text{rms}}$ KS statistic below that of the uncalibrated Sionna input, with values of $0.210$, $0.224$, and $0.0845$ at $92$, $93$, and $94$~GHz, respectively. The identity twin yields larger KS values of $0.668$, $0.752$, and $0.911$ at the same carriers. The only exception among the baselines occurs at $93$~GHz, where FCNN gives a slightly lower KS value of $0.201$. However, FCNN's lower 93~GHz KS is accompanied by worse PDP and CFR-magnitude errors and therefore does not represent the same overall calibration quality. Because the $92$--$94$~GHz datasets are not used for training, model selection, or carrier-specific tuning, these results provide direct evidence of zero-shot cross-frequency generalization to three adjacent W-band carriers. They also constitute an unseen-carrier stress test. The test remains limited to the measured $92$--$95$~GHz span and the present static LOS scene. Denser multipath, non-line-of-sight (NLOS) propagation, and geometry changes require corresponding RT and measured datasets.

Three complementary mechanisms support the observed stability of SC-SLA under carrier shift, and the ablation in Section~\ref{sec:ablation} isolates each of them. The $\tau_{\text{rms}}$ quantile loss aligns the upper tail of the $\tau_{\text{rms}}$ distribution, rather than only its mean and variance. The $\tanh$-bounded residual head keeps the translated CFR within the shared normalized channel range, limiting unstable extrapolation at unseen carriers. The sub-band PDP loss further encourages preservation of delay structure over local frequency regions, making the correction less dependent on the exact $95$~GHz training carrier.

\subsection{Inference Scalability}
\label{sec:scalability}

Cross-frequency transfer avoids carrier-specific retraining. However, the translated CFRs must also be generated at the scale of an RT dataset. Table~\ref{tab:scalability} therefore reports post-RT calibration time for subsets of the $50{,}000$-sample 95~GHz Sionna dataset. The deployed $G_{S\to R}$ contains $2.62$ million parameters and sustains approximately $5.1\times10^3$ CFRs/s across the evaluated dataset sizes. It processes the complete set in $9.75\pm0.09$~s, corresponding to an amortized inference time of $0.195$~ms per CFR. Joint training incorporates two generators with $5.24$ million parameters in total, whereas inference retains only $G_{S\to R}$. The timing includes host-to-device transfer, one generator pass, and output return. The reported timing excludes RT generation, model loading, normalization, and metric computation.

\begin{table}[t]
\caption{FP32 inference scalability of $G_{S\to R}$ on an RTX 4000 Ada GPU with batch size 64 and 1,500 complex64 subcarriers per CFR. Runtime is the mean $\pm$ standard deviation over five runs.}
\label{tab:scalability}
\centering
\footnotesize
\setlength{\tabcolsep}{2.0pt}
\renewcommand{\arraystretch}{1.08}
\begin{tabular}{@{}rrrrr@{}}
\toprule
\textbf{CFRs} &
\shortstack{\textbf{Input}\\\textbf{(MB)}} &
\shortstack{\textbf{Parameters}\\\textbf{($10^6$)}} &
\shortstack{\textbf{Runtime}\\\textbf{(s)}} &
\shortstack{\textbf{Throughput}\\\textbf{(CFRs/s)}} \\
\midrule
$2{,}000$  & $24$  & $2.618$ & $0.380\pm0.001$ & $5{,}257$ \\
$10{,}000$ & $120$ & $2.618$ & $1.910\pm0.005$ & $5{,}235$ \\
$50{,}000$ & $600$ & $2.618$ & $9.751\pm0.085$ & $5{,}127$ \\
\bottomrule
\end{tabular}
\end{table}

\subsection{Sensitivity to Measured Training Data}
\label{sec:data_sensitivity}

Practical calibration also depends on how much measured CSI is available for training. We therefore retrain SC-SLA with $11{,}250$, $22{,}500$, and $33{,}750$ measured 95~GHz CFRs, corresponding to nested $25\%$, $50\%$, and $75\%$ subsets of the measured training set. All $45{,}000$ simulated training CFRs and the fixed $5{,}000$-sample reserved split remain unchanged. Each variant is initialized independently and trained once with seed $1234$ and the original 300-epoch schedule, and every epoch contains 150 optimizer updates, so all fractions receive the same number of gradient steps. The loss weights and checkpoint-selection criterion are unchanged, the normalization parameters remain those estimated from the complete 95~GHz dataset, and the $100\%$ row reuses the full-data checkpoint. This experiment therefore isolates the number of measured CFRs entering optimization rather than the total measurement-acquisition requirement.

Table~\ref{tab:data_sensitivity} reports the results. At the 95~GHz training carrier, the reduced-data models attain $\tau_{\text{rms}}$ KS values of $0.0250$--$0.0350$, below the $0.0495$ of the full-data model, and CFR-magnitude NMSE of $0.1008$--$0.1067$, within $0.0069$ of the full-data $0.1077$. PDP NMSE is the most sensitive in-domain metric at $1.34$--$3.49$ times the full-data value of $3.02\times10^{-5}$, although even its worst case remains more than $50$ times below the uncalibrated error.

\begin{table}[!t]
\caption{Measured-data sensitivity at 95~GHz and under zero-shot transfer to 92--94~GHz. Lower values indicate better agreement. The $100\%$ rows reproduce the full-data SC-SLA results.}
\label{tab:data_sensitivity}
\centering
\footnotesize
\setlength{\tabcolsep}{2.5pt}
\renewcommand{\arraystretch}{1.10}
\begin{tabular}{@{}rrrrr@{}}
\toprule
\multicolumn{5}{c}{\textbf{(a) In-domain evaluation at 95~GHz}} \\
\cmidrule(lr){1-5}
\shortstack{\textbf{Measured}\\\textbf{CFRs}} &
\shortstack{\textbf{Fraction}\\\textbf{(\%)}} &
\shortstack{\textbf{$\tau_{\text{rms}}$}\\\textbf{KS}} &
\shortstack{\textbf{PDP}\\\textbf{NMSE}} &
\shortstack{\textbf{CFR}\\\textbf{magnitude}\\\textbf{NMSE}} \\
\midrule
$11{,}250$ & $25$ & $0.0350$ & $1.05\times10^{-4}$ & $0.1055$ \\
$22{,}500$ & $50$ & $0.0295$ & $4.05\times10^{-5}$ & $0.1008$ \\
$33{,}750$ & $75$ & $0.0250$ & $6.73\times10^{-5}$ & $0.1067$ \\
\midrule
$45{,}000$ & $100$ & $0.0495$ & $3.02\times10^{-5}$ & $0.1077$ \\
\midrule
\multicolumn{5}{c}{\textbf{(b) Zero-shot mean at 92--94~GHz}} \\
\cmidrule(lr){1-5}
\shortstack{\textbf{Measured}\\\textbf{CFRs}} &
\shortstack{\textbf{Fraction}\\\textbf{(\%)}} &
\shortstack{\textbf{$\tau_{\text{rms}}$}\\\textbf{KS}} &
\shortstack{\textbf{PDP}\\\textbf{NMSE}} &
\shortstack{\textbf{CFR}\\\textbf{magnitude}\\\textbf{NMSE}} \\
\midrule
$11{,}250$ & $25$ & $0.1935$ & $2.13\times10^{-3}$ & $0.0726$ \\
$22{,}500$ & $50$ & $0.2112$ & $5.36\times10^{-3}$ & $0.0702$ \\
$33{,}750$ & $75$ & $0.2902$ & $4.46\times10^{-3}$ & $0.0683$ \\
\midrule
$45{,}000$ & $100$ & $0.1728$ & $5.29\times10^{-3}$ & $0.0708$ \\
\bottomrule
\end{tabular}
\end{table}

Cross-frequency behavior is less uniform. The mean $\tau_{\text{rms}}$ KS at 92--94~GHz is $0.1935$, $0.2112$, and $0.2902$ for the $25\%$, $50\%$, and $75\%$ models, against $0.1728$ for the full-data model and $0.7765$ for the uncalibrated twin. Mean PDP NMSE spans $2.13\times10^{-3}$--$5.36\times10^{-3}$, compared with $5.29\times10^{-3}$, and mean CFR-magnitude NMSE stays within $0.0025$ of the full-data $0.0708$. The $75\%$ model yields the largest mean KS but the lowest mean CFR-magnitude NMSE, so sensitivity to the measured-data fraction depends on which channel statistic is evaluated. Under the fourfold reduction, the $25\%$ model still holds in-domain KS below the full-data value and differs from it by $0.0207$ in unseen-carrier mean KS. Each fraction is nevertheless represented by a single run, so these results do not establish a monotonic data-scaling relationship or across-seed uncertainty.

\subsection{Loss Function Ablation Study}
\label{sec:ablation}

We retrain six SC-SLA variants to isolate the role of the main loss and architecture components. Each variant changes one ingredient by removing one channel-statistics loss, removing all four delay-domain losses jointly (\emph{only mag}), removing cycle and identity regularization, or replacing the $\tanh$-bounded head with a linear residual head, $\hat{\mathbf{x}}=\mathbf{x}+\Delta\mathbf{x}$. All variants use the same seed, training schedule, and checkpoint-selection rule as the full model. Because the study uses a single seed, we focus on clear multi-fold differences and avoid overinterpreting small KS changes below roughly $0.02$.

The ablation results demonstrate that the delay and frequency-domain objectives are complementary. Removing the $\tau_{\text{rms}}$ quantile loss increases the in-domain KS statistic by $3.6\times$ ($0.0495 \rightarrow 0.1775$) and worsens the $94$~GHz KS by $2.6\times$, while PDP and magnitude errors remain nearly unchanged. These multi-fold KS increases indicate that the quantile loss primarily controls the upper tail of the $\tau_{\text{rms}}$ distribution. In contrast, the \emph{only mag} variant gives the lowest CFR-magnitude NMSE in Fig.~\ref{fig:abl_indomain} ($0.0999$), but leaves the delay domain largely uncorrected, with KS $0.694$. Removing the magnitude loss produces the opposite failure mode. The delay-domain terms remain active, but the spectral profile drifts and the CFR-magnitude NMSE rises to $0.1608$, worse than the uncalibrated input.

The architecture ablations clarify the source of sample-level input--output coupling. A linear residual head fits the training carrier reasonably well, with in-domain KS $0.0425$, but generalizes poorly. The $94$~GHz KS more than doubles from $0.0845$ to $0.1875$, and the in-domain PDP NMSE increases by $5.7\times$. Thus, the $\tanh$ head mainly stabilizes the correction under carrier shift rather than improving the 95~GHz fit alone. Removing the cycle and identity losses does not cause collapse. On $512$ held-out inputs, the mean input--output correlation remains $0.82$, compared with $0.75$ for the full model. These correlation values indicate that, in the present static single-link setting, the bounded residual form provides most of the input--output coupling. We retain cycle and identity regularization because it is conservative and may become more important when geometry varies across samples.

The sub-band PDP loss contributes mainly to cross-frequency robustness. Removing it slightly improves some in-domain metrics, but worsens the $94$~GHz KS statistic by $43\%$ ($0.0845 \rightarrow 0.1205$) and also degrades KS at $92$ and $93$~GHz. Across Figs.~\ref{fig:abl_indomain} and~\ref{fig:abl_crossfreq}, no ablated model is uniformly superior. Each variant improves one metric at the cost of another. The full configuration is therefore not the per-metric optimum, but it provides the most balanced operating point across $\tau_{\text{rms}}$, PDP shape, spectral magnitude, and carrier transfer.
\section{Conclusion}
\label{sec:conclusion}

In this paper, we presented SC-SLA, a non-adversarial, cycle-consistent framework that calibrates a ray-traced W-band CT against unpaired testbed measurements by aligning receiver-relevant OFDM channel statistics. These statistics include the PDP, sub-band PDP, $\tau_{\text{rms}}$ distribution, and per-subcarrier CFR magnitude. On $2{,}000$ held-out captures at $95$~GHz, SC-SLA reduced the $\tau_{\text{rms}}$ KS statistic of the uncalibrated twin from $0.86$ to $0.0495$, improved on the strongest of four supervised baselines by $44\%$, and achieved the lowest PDP and CFR-magnitude NMSE among all evaluated models. The same checkpoint transferred zero-shot to $92$--$94$~GHz, providing evidence of cross-frequency calibration over the measured $3$~GHz span using training measurements from 95~GHz only. A fourfold reduction of the measured training set to $11{,}250$ CFRs preserved calibration quality under a fixed seed, yielding KS $0.0350$ at 95~GHz and mean KS $0.1935$ across the three unseen carriers. The ablation study demonstrated that the $\tau_{\text{rms}}$ quantile loss governs the upper tail of the $\tau_{\text{rms}}$ distribution, the delay- and frequency-domain objectives are jointly necessary, and the $\tanh$-bounded residual head stabilizes the correction under carrier shift. The deployed generator calibrates $50{,}000$ CFRs in $9.75\pm0.09$~s, corresponding to approximately $5.1\times10^3$ CFRs/s. Future work could repeat the sensitivity and ablation studies across seeds, quantify sensitivity to the impairment parameters, extend the captures beyond the present \SI{50}{\MHz} sampling bandwidth, and address NLOS, geometry-varying, and multi-antenna deployments.

\bibliographystyle{IEEEtran}
\bibliography{references}
\end{document}